\documentclass[notitlepage,aps,12pt,showpacs,showkeys,tightenlines,nofootinbib]{revtex4}

\usepackage{amsmath}
\usepackage{amssymb,amsfonts}
\usepackage{stmaryrd}

\newcommand{\comment}[1]{}

\newcommand{\ie}{\textit{i.e.}}
\newcommand{\etal}{\textit{et~al.}}

\newcommand{\tsfrac}[2]{{\textstyle\frac{#1}{#2}}}

\renewcommand{\i}{\mathrm{i}}

\usepackage[dvips]{epsfig}
\usepackage{psfrag}

\graphicspath{{Figures/}{Figures_lo/}}

\begin{document}

\title{Topological Chaos in Spatially Periodic Mixers}
\author{Matthew D. Finn}
\author{Jean-Luc Thiffeault}
\email{jeanluc@imperial.ac.uk}
\author{Emmanuelle Gouillart}
\affiliation{Department of Mathematics, Imperial College
   London, SW7 2AZ, United Kingdom}
\date{\today}

\keywords{chaotic advection, topological chaos}
\pacs{47.52.+j, 05.45.-a}

\begin{abstract}
Topologically chaotic fluid advection is examined in two-dimensional flows
with either or both directions spatially periodic. Topological chaos is
created by driving flow with moving stirrers whose trajectories are chosen to
form various braids.  For spatially periodic flows, in addition to the usual
stirrer-exchange braiding motions, there are additional
topologically-nontrivial motions corresponding to stirrers traversing the
periodic directions. This leads to a study of the braid group on the cylinder
and the torus. Methods for finding topological entropy lower bounds for such
flows are examined.  These bounds are then compared to numerical stirring
simulations of Stokes flow to evaluate their sharpness.  The sine flow is also
examined from a topological perspective.
\end{abstract}

\maketitle

\section{Introduction}

\begin{figure}
\begin{center}
\epsfig{file=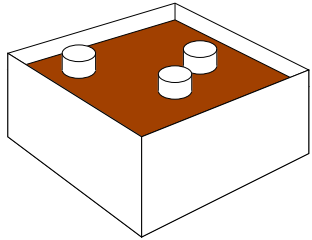,scale=0.35}
\end{center}
\caption{The prototype batch stirring device. This bounded mixer may be made
  spatially periodic by removing either or both pairs of opposite walls and
  identifying the edges of the fluid. When this is done, extra stirrer motions
  are possible that traverse the periodic directions.  Vertical flow effects
  are neglected.}
\label{fig:mixer}
\end{figure}

We study topological chaos in regimes where two-dimensional flow is driven by
moving stirrers, such as in a prototypical batch stirring
device~\cite{Boyland2000,Finn2003a} illustrated in Figure~\ref{fig:mixer}. We
consider the effect of extra stirrer motions that are possible when the fluid
domain has spatial periodicity. For the mixer in Figure~\ref{fig:mixer}, we
imagine removing one or both pairs of opposite walls and identifying these
edges to make a periodic domain. Our study is motivated by the many model
flows examined in the literature that have spatial periodicity, such as the
ubiquitous sine flow~\cite{Pierrehumbert1994}.  Some laboratory experiments
are also designed to have spatial periodicity in one
direction~\cite{Paoletti2005}, and the square patterns formed in thermal
convection exhibit periodicity in two directions~\cite{Chandra}.  There are
also signal analysis applications of braid groups~\cite{Amon2004} where
measured quantities can be periodic.

Topological chaos (TC) can be used to design mixers with a robust built-in
stirring quality that depends only on the relative motions of the stirrers and
is independent of fluid properties such as compressibility or
viscosity~\cite{Boyland2000}. This mixing quality can be guaranteed because
stirrers are topological obstacles to flow: if their trajectories form an
intertwined braiding pattern, then by continuity material lines in the fluid
must also be braided.  If the stirrer motion traces a braid with positive
topological entropy, then typical material lines in the mixing region of the
fluid will grow exponentially.  Roughly speaking, this means that the braiding
motion of the stirrers guarantees a minimum amount of chaos in the flow in a
region around the stirrers, and chaos leads to efficient fluid
mixing~\cite{Aref1984}.  In the two-dimensional flows studied here, the
topological entropy is related to the time-asymptotic exponential growth rate
of material lines.  Since the length of material lines increases rapidly, they
must fold back and forth within the bounded mixing region and lead to the
familiar ``striations'' observed in typical mixing flows.

The properties of a braid are linked to Thurston--Nielsen
theory~\cite{Boyland1994,Thurston1988}.  Thurston proved that a
two-dimensional homeomorphism (here, the map giving the motion of fluid
particles in one period) can be decomposed into regions that are either
reducible, finite-order, or isotopic to a pseudo-Anosov map.  Braids are a
convenient language to describe the \emph{isotopy classes} of these
homeomorphisms.  For our purposes, the third type is most important:
pseudo-Anosov maps have strong chaotic properties and typically lead to good
mixing.  If the braid is \emph{irreducible} with positive entropy, then to it
must correspond a region of the flow that is a single chaotic component.  In
fluid mixing parlance, this region will be well-mixed (though it could be very
small or even have zero measure).

To guarantee positive topological entropy in a nonperiodic domain, it can be
shown that at least three trajectories must be involved in a braid. However,
this does not imply that mixers must have three or more stirrers to produce
TC~\cite{Gouillart2006,Thiffeault2005}. This is because we can consider
trajectories of anything that acts as a topological obstacle to material
lines: in a two-dimensional flow this could be a fluid particle itself, or an
entire island of regularity. Thus, all mixers generally produce TC, since it
is easy to find sets of periodic trajectories that braid. This means that
topological ideas can also be used to understand mixing, even for flows with
fewer than three stirrers~\cite{Gouillart2006,Thiffeault2005}. It is also
instructive to note that all chaos in time-periodic two-dimensional flows is
topological, in the sense that the topological entropy of a flow can be
arbitrarily well-approximated by considering the braiding motion of a
sufficiently high-order periodic orbit~\cite{Katok1980,Boyland1994}.

We begin in Section~\ref{sec:bounded} by summarising current techniques that
have been applied to bounded flows (or spatially-nonperiodic flows on the
infinite plane).  Then in Section~\ref{sec:periodic} we describe ways of
extending these methods for spatially periodic flows, examining in turn the
case of single- and double-periodicity. In Section~\ref{sec:numerics} we
determine the sharpness of theoretical topological entropy estimates compared
to entropies for model Stokes flows.  We apply our techniques to the sine flow
in Section~\ref{sec:sineflow}, where the ``stirrers'' are given by periodic
orbits.  Finally, we close with a discussion of our results in
Section~\ref{sec:discussion}.

\section{Braiding in a Nonperiodic Domain}
\label{sec:bounded}

In this section we describe how motion of stirrers around each other in a
nonperiodic domain (\ie, a surface of genus zero such as a bounded domain or
the infinite plane) may be described as a braid.  We then review how this
braid may be used to produce a rigorous lower bound on the topological entropy
of the flow.  We consider two-dimensional flows, where the topological entropy
is the time-asymptotic growth rate of material
lines~\cite{Franks1988,Newhouse1993}.  (It is the supremum over all possible
smooth material lines, but in practice we find that the growth rate rapidly
becomes independent of the choice of initial material line, except for
pathologically bad choices.)

\subsection{Artin's Braid Group}

In flows without spatial periodicity, braiding is conventionally characterised
by recording exchanges of position of adjacent
stirrers~\cite{Boyland2000}. Stirrer positions are projected onto an axis (we
have chosen the $x$-axis in what follows) and ordered from $1$ to $n$ along
this line.  If stirrer $i$ and $i+1$ exchange order in a clockwise direction,
we assign to this motion the braid group generator $\sigma_i$; an
anti-clockwise crossing is labelled $\sigma_i^{-1}$.  The numbers $1$ to $n$
always refer to the relative position of the stirrers along the $x$-axis, and
do not always label the same stirrer.  The procedure of associating
trajectories with braid generators is explained in detail in
Refs.~\cite{Thiffeault2005,Vikhansky2003a}.  Any motion of stirrers then
corresponds to an ordered sequence of generators: this sequence is an element
of Artin's braid group~\cite{Birman1975,Murasugi} on $n$ strands. This group
contains all braids generated by sequences of $\sigma_1,\ldots,\sigma_{n-1}$
and their inverses.  The group operation is the catenation of braids, and the
inverse is found by reversing the sequence and inverting the generators.  The
identity of the braid group is simply $n$ parallel strands.  We use the
convention that braid generators occur from left to right, so that in the
braid word $\sigma_1\sigma_2$ the generator $\sigma_1$ occurs first.

Some braids are topologically equivalent because they can be continuously
deformed into each other. In fact, equivalence of braids can be established
using just the group relations
\begin{subequations}
\begin{alignat}{2}
\sigma_i \sigma_{j} \sigma_i &= \sigma_{j} \sigma_{i} \sigma_{j} \qquad
&\mbox{if} \qquad |i-j| &= 1, \label{eq:121} \\
\sigma_i \sigma_j &= \sigma_j \sigma_i \qquad &\mbox{if} \qquad |i-j| &\ge 2,
\label{eq:13}
\end{alignat}
\end{subequations}
which together form a presentation for the group (a minimal set of relations
describing the group).  Relation \eqref{eq:121} implies that particular
sequences of three braid elements are equivalent, and is obvious by inspecting
a diagram of the braid~\cite{Murasugi}. Relation \eqref{eq:13} dictates that
any two braid events involving different pairs of stirrers can be commuted.

The braid word describing the relative motions of the stirrers encodes a great
deal of information about the fluid flow. One such piece of information is the
topological entropy of the braid, which is a lower bound on the topological
entropy of the flow. Two methods of calculating this lower bound have been
considered: matrix representations of the braid group
\cite{Boyland2000,Kolev1989}, and train-tracks~\cite{Bestvina1995}.  For three
stirrers, the two are equivalent; for more than three stirrers the matrix
method only gives a lower bound on the topological entropy of the braid.
Hence in this paper we shall use exclusively train-tracks to compute the
topological entropy of a braid.  This is discussed in the following sections.

\subsection{Train-tracks}
\label{sec:tracks}

A complicated but robust method for computing braid topological entropies is
the Bestvina--Handel train-track algorithm \cite{Bestvina1995}.  This
algorithm requires the construction of a special invariant graph for each
braid.  Computerised implementations of this algorithm exist that take a word
in generators of Artin's braid group as an
input~\cite{Brinkmann2000,HallTrain}.  These are not currently well-suited to
the periodic domains that we consider here, though they could be adapted for
this purpose.  Here we shall be dealing with braids that are simple enough
that we will not need the Bestvina--Handel algorithm.  Rather, we will use the
method that Thurston calls `iterate and guess,' which is based on the fact
that under iteration almost any curve converges to the train-track (for
pseudo-Anosov braids).

A train-track consists of a set of edges, joined together at junctions. For
the braids studied in this paper the following types of edge are needed:
\begin{enumerate}
\item Edges that wrap tightly around the stirrers in a closed loop starting
and finishing at the same junction;
\item Edges that join the junctions associated with different stirrers.
\end{enumerate}
The edges of a train-track graph are deformed under the action of the braid,
and we assume the edges are like elastic bands, so that they stretch by the
smallest amount possible without crossing through any of the stirrers.

Type 1 edges, since they are wrapped tightly around the stirrers, are not
stretched under the action of the braid. However, under the action of the
braid the stirrers may be permuted, so Type 1 edges are permuted.  Type 2
edges are non-trivially deformed under the braid and may become wrapped around
several stirrers.

A graph is said to be invariant if under the action of the braid all edges are
mapped onto edge-paths of the original graph.  In general, Type 2 edges are
mapped onto a edge-paths made up of Type 1 and Type 2 edges.  Under the action
of the flow, a material line representing edges from the train-track must be
stretched at least as much as the train-track itself. However, in general it
is stretched even more due to the presence of additional periodic orbits which
act as obstacles to the flow (more on this in Section~\ref{sec:discussion}).

To summarise the technique so far: A graph of Type 1 and 2 edges is drawn
between and around the stirrers.  Then the stirrers are made to undergo the
braid operation under investigation, and the deformation of the graph is
tracked as if its edges were rubber bands.  The goal is to find a graph that
is invariant under the action of the braid---that is, the edges may stretch
and fold, but they all end up at positions where there was initially an
edge-path.  Of course, the edges get permuted and stretched, so a single edge
may be mapped to a path that traverses the original edges several times.  We
call the invariant graph a train-track for the braid.  (In order to be a
genuine train-track, a graph must have a special non-cancellation property
that we discuss below.)

Once the train-track graph has been found, the stretching in the train-track
may be measured by constructing a transition matrix describing for each edge
the edge-path into which it is mapped.  The logarithm of the largest
eigenvalue of this matrix is the topological entropy of the braid.  The
largest eigenvalue is used because it is assumed that the braid operation is
repeated a large number of times, so the graph converges to an invariant
pattern corresponding to the eigenvector of the largest eigenvalue.  The
entries in the eigenvector describe the relative number of edges that
accumulate under repeated action of the braid.  The eigenvalue then describes
the growth of the length of this invariant pattern at each application of the
braid.

\begin{figure}
\begin{center}
\psfrag{edgea}{$\!a$}
\psfrag{edgeb}{$\!b$}
\psfrag{edge1}{$1$}
\psfrag{edge2}{$2$}
\psfrag{edge3}{$3$}
\psfrag{(a)}{(a)}
\psfrag{(b)}{(b)}
\epsfig{file=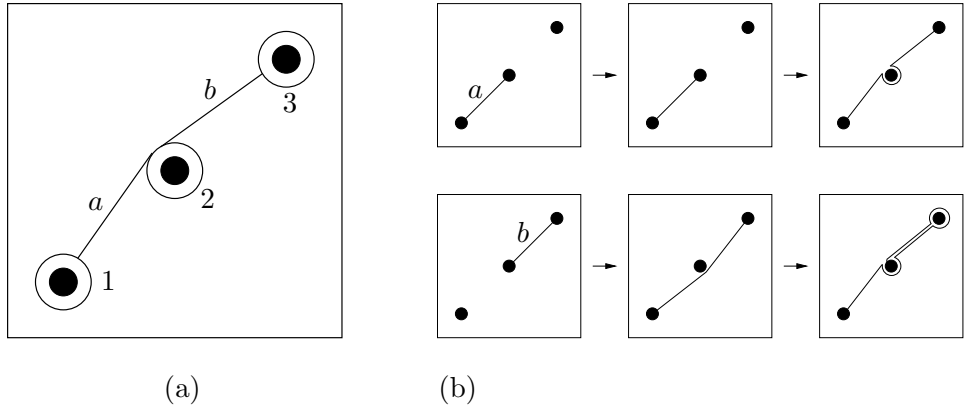,scale=0.5}
\end{center}
\caption{(a) The train-track for the braid $\sigma_1 \sigma_2^{-1}$; (b) The
  deformation of edges $a$ and $b$ under the action of the braid.}
\label{fig:btrack}
\end{figure}

As an example we consider the well-studied pigtail braid $\sigma_1
\sigma_2^{-1}$ with three stirrers, with train-track illustrated in
Figure~\ref{fig:btrack}(a). It consists of three Type 1 edges, labelled 1, 2
and 3, that pass around each of the stirrers, and two Type 2 edges, labelled
$a$ and $b$, that join adjacent stirrers.
The two sequences shown in Figure~\ref{fig:btrack}(b) illustrate the fate of
the edges $a$ and $b$ under $\sigma_1$ followed by $\sigma_2^{-1}$. Up to a
deformation that does not change the topology, the edges of the train track are
mapped to the following edge-paths:
\begin{equation}
 a \mapsto a2b, \quad
 b \mapsto a2b3b, \quad
 1 \mapsto 3, \quad
 2 \mapsto 1, \quad
 3 \mapsto 2.
\label{eq:transmap}
\end{equation}
An important feature to note in this example is that ultimately the edges
$a$ and $b$ stretch from the first stirrer, {\it underneath} the second one,
to the third one; hence the resultant edge-paths include a tour around edge
$2$.

In every train-track all edges must be mapped to an edge-path that has
alternately Type 2 edges (denoted by letters) and Type 1 edges (denoted by
numbers). This ensures that two or more adjoining Type 2 edges do not collapse
back to a shorter edge-path under repeated action of the braid.  (This is the
non-cancellation property alluded to above.)  With no possibility of such
cancellations a transition matrix can be written down by recording the number
of edges in each edge-path.  This matrix can then be used to determine the
evolution of an arbitrary edge-path under any number of subsequent
applications of the braid.

The transition matrix for the braid word $\sigma_1 \sigma_2^{-1}$ is found
from \eqref{eq:transmap} to be
\begin{equation}
M = \left[
\begin{array}{cc|ccc}
1 & 1 & 0 & 0 & 0 \\
1 & 2 & 0 & 0 & 0 \\ \hline
0 & 0 & 0 & 1 & 0 \\
1 & 1 & 0 & 0 & 1 \\
0 & 1 & 1 & 0 & 0 \\
\end{array}
\right]%
,
\label{eq:transmx}
\end{equation}%
where the element~$M_{ij}$ represents the number of occurences of the $i$th
edge in the image of the $j$th edge, and the indices~$i$ and $j$ take
values~$(a,b,1,2,3)$, in that order.  Let the vector $(n_a,n_b,n_1,n_2,n_3)^T$
contain the number of edges of each type in a path. Under the action of the
braid these will be mapped to a longer path with edges given by $M \cdot
(n_a,n_b,n_1,n_2,n_3)^T$.

The topological entropy of the flow is given in two dimensions by the
time-asymptotic growth rate of material lines (maximised over all possible
smooth lines).  Since the train-track could be considered as a material line,
the growth rate of the flow must be at least as large as the growth rate of
the train-track. This is given by the spectral radius (the magnitude of the
largest eigenvalue) of the transition matrix, because for a large number of
repeated applications of the braid the coiling will become dominated by the
largest eigenvector of the transition matrix.

For the pigtail braid, the matrix~\eqref{eq:transmx} has a spectral radius
of~$2.62$. Hence the topological entropy of the braid $\sigma_1 \sigma_2^{-1}$
is~\hbox{$\log 2.62 \approx 0.96$}. Any time-periodic flow which has material
obstacles executing this braiding motion necessarily has a topological entropy
of at least~$0.96$.

Note that since Type 1 edges are wrapped tightly around stirrers, and are only
permuted amongst themselves under the braid, they do not
contribute to growth of material lines.  It is thus usual to write the
transition matrix only for Type 2 edges. In the present example this is the
upper-left block indicated in~\eqref{eq:transmx}.

\section{Braiding on the Cylinder and Torus}
\label{sec:periodic}

In this section we study the braid group on the cylinder and the torus.
Because of the possibility of stirrers going around the periodic direction(s),
these groups are different to Artin's braid group discussed in
Section~\ref{sec:bounded}.  Note that Thurston--Nielsen theory has been
considered before on the annulus and torus, see for example~\cite{Boyland1988}
and~\cite{Llibre1991}.  Here we focus on a braid description of stirrer motion
on these surfaces, using for the most part the notation of
Birman~\cite{Birman1969}.

\subsection{Braid Groups on Periodic Domains}

\begin{figure}
\begin{center}
\psfrag{xaxis}{$x$}
\psfrag{yaxis}{$y$}
\psfrag{tau3}{$\tau_3$}
\psfrag{rho4}{$\rho_4$}
\psfrag{sigma1}{$\sigma_1$}
\psfrag{rod1}{$1$}
\psfrag{rod2}{$2$}
\psfrag{rod3}{$3$}
\psfrag{rod4}{$4$}
\psfrag{rodn}{$n$}
\epsfig{file=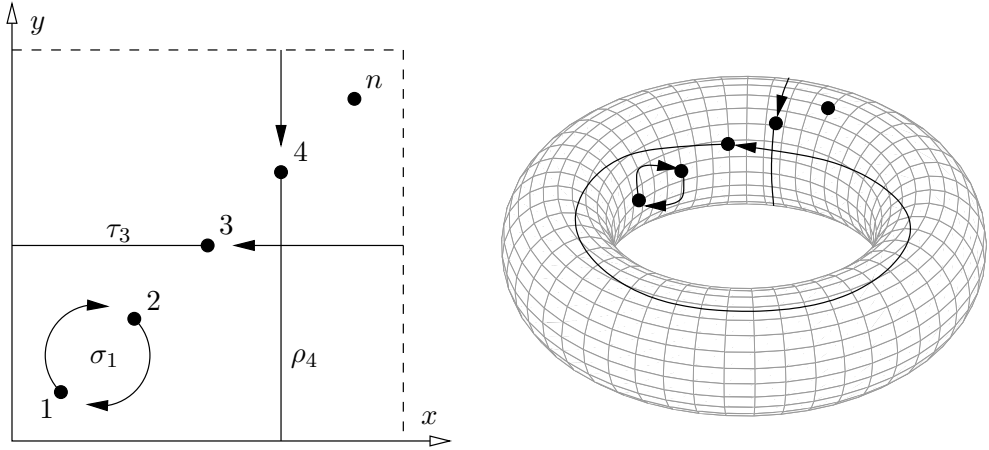,height=6cm}
\end{center}
\caption{Braiding operations on the cylinder and torus. In addition to the
  usual $\sigma_i$, corresponding to interchanging adjacent stirrers, there
  are new braiding operations corresponding to stirrers making a tour around
  one or both periodic directions. For domains that are periodic in only one
  direction (cylinder) we allow the generators~$\tau_i$. For doubly-periodic
  domains (torus) we also consider the generators~$\rho_i$.}
\label{fig:zplane}
\end{figure}

For a periodic domain, in addition to the braid elements $\sigma_i$
representing the exchange of stirrer positions, we introduce extra motions
corresponding to stirrers making tours of the periodic directions, as
illustrated in Figure \ref{fig:zplane}.  In a singly-periodic domain we allow
the extra braid motions $\tau_i$ corresponding to the $i$th stirrer touring
the periodic direction.  The singly-periodic case represents flow on a
(possibly infinite) cylinder.  In this paper we choose the periodicity of the
cylinder to be in the $x$-direction without loss of generality.  For a
doubly-periodic flow we allow the extra braid motions $\tau_i$ and
$\rho_i$---the additional motions $\rho_i$ correspond to the $i$th stirrer
touring the second periodic direction. The doubly-periodic case can be thought
of as flow on a torus.

The braiding motions $\tau_i$ and $\rho_i$ are not independent of the
$\sigma_i$ nor each other, since some braids can be continuously deformed into
each other.  The braid group relations for an arbitrary number of stirrers
have been stated by Birman~\cite{Birman1969}.  As an example, for the braid
group on the torus with two stirrers, the one relation required to determine
equivalence of braids is~\cite{Birman1969}
\begin{equation}
\sigma_1^2 = \rho_1^{-1} \tau_2 \rho_1 \tau_2^{-1}.
\label{eq:toruspres}
\end{equation}
The relation~\eqref{eq:toruspres} provides a presentation for the two-strand
braid group on the torus, with generators~$\{\sigma_1,\rho_1,\tau_2\}$.  The
other two operations can be written~$\rho_2=\sigma_1\rho_1\sigma_1$,
$\tau_1=\sigma_1\tau_2\sigma_1$.

The ideas developed in the following section
apply for an arbitrary number of stirrers, but we shall illustrate flow with
just two stirrers.  We will see that is possible to create topological chaos
on a periodic domain with fewer than three stirrers.

\subsection{Representation for the Cylinder}
\label{sec:cylinder}

For the special case of the cylinder, we consider braids formed from
$\sigma_i$ and $\tau_i$ generators only, and ignore $\rho_i$ for the moment.
The dynamics on a singly-periodic domain is only a little more complicated
than the dynamics in a bounded domain.  This is because, as pointed out in
Ref.~\cite{Boyland2003}, there exists a conformal mapping from the periodic
strip (\ie, the cylinder) to an annulus, and the topological entropy of a flow
is preserved under this conformal mapping.

\begin{figure}
\begin{center}
\psfrag{rod1}{$1$}
\psfrag{rod2}{\raisebox{-2pt}{$2$}}
\psfrag{rod3}{\raisebox{-3pt}{$3$}}
\psfrag{rod4}{$4$}
\psfrag{rodn}{$n$}
\psfrag{nplus1}{$\!n+1$}
\psfrag{sigma1}{$\sigma_1$}
\psfrag{tau3}{$\tau_3$}
\epsfig{file=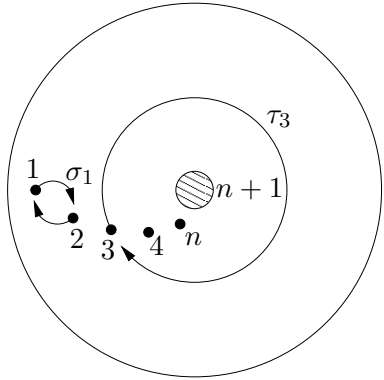,height=5cm}
\end{center}
\caption{Plot of the complex $w$-plane after making the conformal mapping
  $w=\exp(2\pi\i z)$ from a singly-periodic domain with $n$ stirrers to an
  annular domain. The central region (shaded) acts as an $(n+1)$th
  stirrer. The braid element $\tau_i$ corresponds to the $i$th stirrer making
  a clockwise tour of all the stirrers $i+1$, $\ldots$, $n+1$, so that
  $\tau_i\mapsto\Sigma_i\ldots\Sigma_n\Sigma_n\ldots\Sigma_i$.}
\label{fig:wplane}
\end{figure}

In Figure \ref{fig:wplane} we show the result of conformally mapping the
square in the $z$-plane (with $z=x+i y$) in Figure \ref{fig:zplane} under the
mapping $w=\exp(2\pi\i z)$ onto an annulus in the $w$-plane. The stirrers
aligned along a diagonal are mapped onto a logarithmic spiral in the
$w$-domain.
The key feature in this $w$-domain is that there is a hole (centered on the
origin) that acts as an $(n+1)$th stirrer. In the case of the $z$-domain
extending to $y\rightarrow+\infty$ this hole shrinks to a point, but it is
still a topological obstacle to the flow. If the $z$-domain includes
$y\rightarrow-\infty$ then the $w$-domain extends to infinity.

Braid elements $\sigma_i$ do not involve the hole. However, the periodic
motions $\tau_i$ make a clockwise tour of the hole (stirrer $n+1$), also
crossing all stirrers $i+1,\ldots,n$ twice (Figure~\ref{fig:wplane}). Thus the
braid group on a singly-periodic domain with $n$ strings, with generators
$\{\sigma_i,\tau_1\}$, can be associated with the braid group in the plane
with $n+1$ strings, with generators~$\{\Sigma_i\}$, if we assign
\begin{equation}
  \sigma_i\mapsto\Sigma_i\,, \quad
  \tau_i\mapsto\Sigma_i\ldots\Sigma_n\Sigma_n\ldots\Sigma_i.
\end{equation}
An important consequence of the hole acting as a stationary stirrer is that
only two proper stirrers are required to guarantee topological chaos.  (On a
nonperiodic domain, at least three stirrers are needed, since the topological
entropy of all braids on two strands is zero~\cite{Boyland2000}.)

We illustrate this for the two-stirrer braid $\tau_1 \sigma_1$ which is
equivalent to the three-stirrer planar braid $\Sigma_1 \Sigma_2 \Sigma_2
\Sigma_1 \Sigma_1$.  We will show in the following section that this braid has
a topological entropy of $1.32$.  Note that this is larger than the entropy
$0.96$ for the same length braid word $\sigma_1 \sigma_2^{-1}$, which was
shown in Ref.~\cite{DAlessandro1999} to be optimal on a nonperiodic domain.
This suggests that periodic boundary conditions contribute significantly in
creating chaos!

\subsection{Train-tracks for the Cylinder and Torus}
\label{eq:ttct}

In principle we can compute the topological entropy of braids on the cylinder
by first applying the conformal transformation defined in
Section~\ref{sec:cylinder} and finding the corresponding braid word on the
annulus.  However, the braid word on the annulus is usually considerably
longer than that on the cylinder, which makes the train-track harder to
compute.  Moreover, for the doubly-periodic case (torus) there is no such
conformal mapping trick, so in any case we will need to find train-tracks on
periodic domains.  Fortunately we can find train-tracks directly for the
cylinder and torus by considering graphs with edges that traverse the periodic
directions.  The periodic train-tracks shown in this paper are determined by
inspection, following Thurston's `iterate and guess' approach, though we
expect that existing computer codes (see {\it e.g.}\ \cite{HallTrain}) could
be extended to automate their construction.

The method works as follows.  Consider stretching an elastic band between two
stirrers, possibly traversing the periodic direction in doing so. Now consider
the fate of the elastic under the action of the braid.  The band will be
stretched around the stirrers with the same topology as the train-track for
the braid (except for possible rare choices of initial condition where the
band undergoes a periodic motion without stretching at all).  If an invariant
train-track exists then the elastic band will tend to align with it as it is
stretched. The reason it must align is that initially the elastic crosses the
train-track a fixed number of times (possibly zero), and by determinism it
must still cross the same number of times even after many applications of the
braid. Now, if the elastic is rapidly stretching but never makes any
additional crossings of the train-track, then it must align and be stretched
parallel to the track.  After a few iterations of the braid it is usually
possible to determine the shape of the required invariant graph by
inspection. It can then be verified that the graph is a proper train-track by
checking that all Type 2 edges are mapped onto edge-paths consisting of
alternating Type 2 and Type 1 edges.

\begin{figure}
\begin{center}
\psfrag{(a)}{(a)}
\psfrag{(b)}{(b)}
\psfrag{(c)}{(c)}
\psfrag{edgea}{$a$}
\psfrag{edgeb}{$b$}
\psfrag{edge1}{$1$}
\psfrag{edge2}{$2$}
\epsfig{file=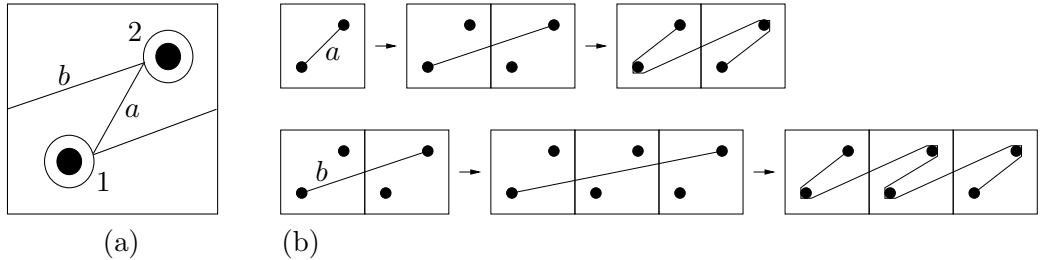,scale=0.44}
\end{center}
\caption{(a) Train-track for the braid $\tau_1 \sigma_1$ on the cylinder; (b)
  The deformation of edges $a$ and $b$ under the action of the braid.}
\label{fig:strack}
\end{figure}

In Figure~\ref{fig:strack} we illustrate construction of a train-track for the
braid $\tau_1 \sigma_1$ studied above. The starting point for this calculation
is to compute the fate of an elastic band $a$ stretched directly between the
two stirrers. Under the action of the braid this is deformed into a `Z'
shape, shown in the top sequence in Figure~\ref{fig:strack}(b). Note that
headaches can be spared by drawing extra copies of the periodic domain as
necessary, rather than attempting to wrap the line around on a single copy of
the domain. The resulting `Z' shape has two edges like the original $a$, and a
new edge---call it $b$---joining stirrer $1$ to $2$ across the periodic
direction.

The bottom sequence in Figure~\ref{fig:strack}(b) shows the evolution of
$b$. The edge $b$ is mapped to a double `Z' shape that spans three copies of
the domain and consists of three $a$ edges and two $b$ edges. Because no
unfamiliar edges are generated we conjecture that $a$ and $b$, plus two Type 1
loops to allow lines to wrap around the stirrers, are all that is required for
the train-track. To confirm that this invariant graph is a proper train-track
we examine the edge-paths produced by the braid, which are
\begin{equation}
a \mapsto a1b2a, \quad
b \mapsto a1b2a1b2a, \quad
1 \mapsto 2, \quad
2 \mapsto 1.
\end{equation}
All the Type 2 edges are mapped to paths of alternating Type 2 and Type 1
edges, so this confirms we have found the required train-track. The transition
matrix for the braid $\tau_1 \sigma_1$ (ignoring Type 1 edges which do not
contribute to stretching) is then
\begin{equation}
\left[
\begin{array}{cc}
2 & 3\\
1 & 2
\end{array}
\right]
\end{equation}
with spectral radius $3.73$, so the braid has a topological entropy of
$1.32$.

\begin{figure}
\begin{center}
\psfrag{edge1}{$1$}
\psfrag{edge2}{$2$}
\psfrag{edgea}{$a$}
\psfrag{edgeb}{$b$}
\psfrag{edgec}{$c$}
\psfrag{edged}{$d$}
\epsfig{file=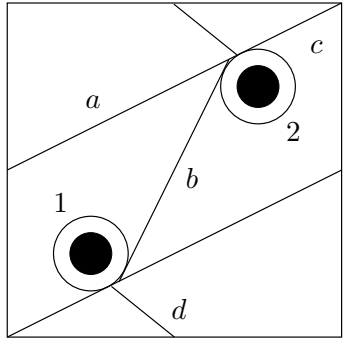,scale=0.5}
\end{center}
\caption{The train-track for the braid $\tau_1 \sigma_1 \rho_1^{-1} \sigma_1$
  on the torus.}
\label{fig:dtrackab}
\end{figure}

\begin{figure}
\begin{center}
\psfrag{edgea}{$a$}
\psfrag{edgeb}{$b$}
\psfrag{edgec}{$c$}
\psfrag{edged}{$d$}
\epsfig{file=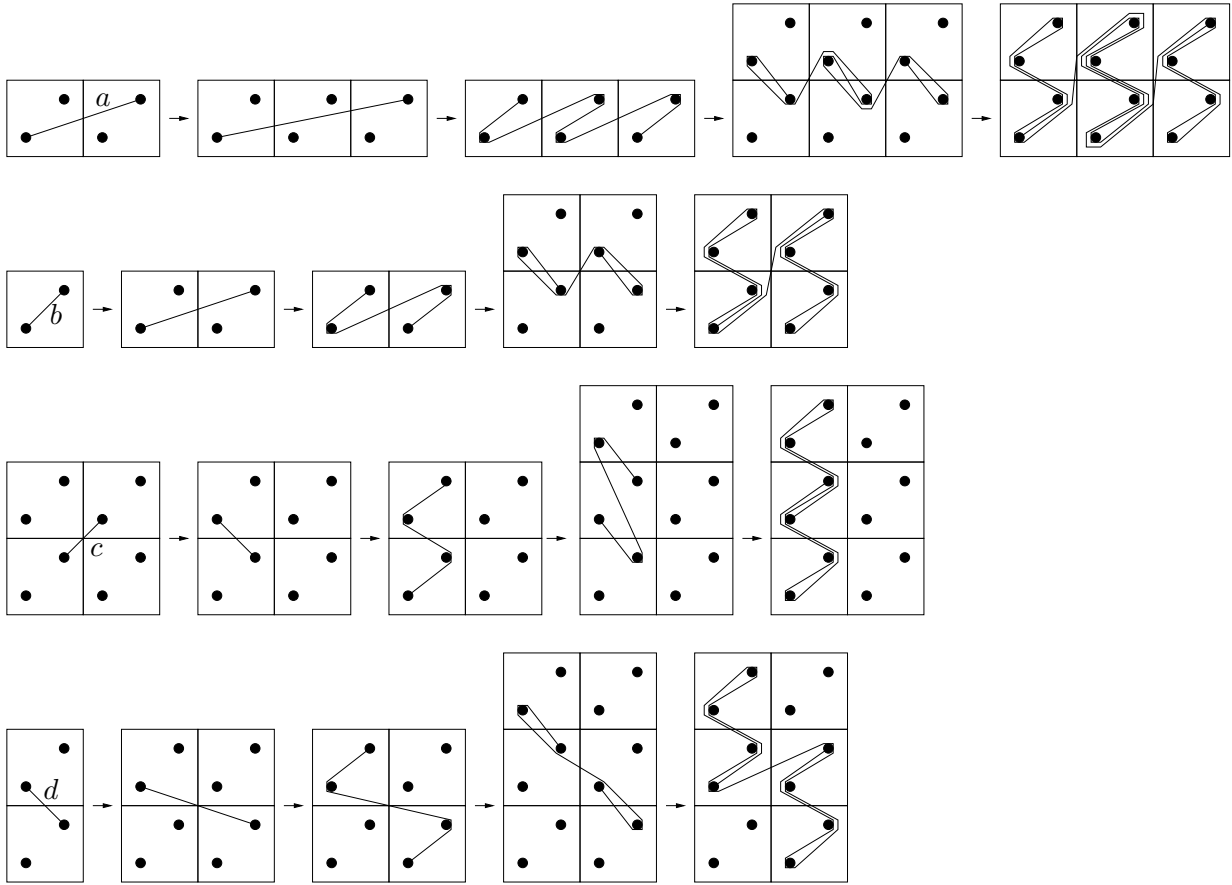,scale=0.4}
\end{center}
\caption{The deformation of edges $a$, $b$, $c$, $d$ in
  Figure~\ref{fig:dtrackab} under the action of the braid $\tau_1 \sigma_1
  \rho_1^{-1} \sigma_1$.}
\label{fig:dtrackedge}
\end{figure}

The same inspection method works for toroidal braids, where we allow the
additional braid elements $\rho_i$ corresponding to stirrers touring the
second periodic direction. We illustrate the construction in
Figures~\ref{fig:dtrackab}--\ref{fig:dtrackedge} for the more complicated
braid $\tau_1 \sigma_1 \rho_1^{-1} \sigma_1$.  Figure~\ref{fig:dtrackab}
depicts the invariant train-track for this braid.%
The action of the braid on the various edges is shown in
Figure~\ref{fig:dtrackedge}.  Our initial `guess' elastic band is labelled
$a$. Under the action of the braid, $a$ is mapped onto three new edges, $b$,
$c$, and $d$ (Figure~\ref{fig:dtrackedge}, top). Edge $b$ is mapped to a path
containing $b$, $c$ and $d$. Edge $c$ is mapped onto $b$ and $d$. Edge $d$ is
mapped onto $a$, $b$ and $d$, and no unfamiliar edges are generated. Hence
$a$, $b$, $c$ and $d$, plus two small loops around the stirrers, may be
checked as a candidate train-track.

Careful inspection reveals edges are mapped to edge-paths as
\begin{align*}
a &\mapsto b2d1b2b1d2b1b2c1b2b1d2b1b2d1b2b1d2b1b2c1b2b1d2b1b2d1b, \\
b &\mapsto b2d1b2b1d2b1b2c1b2b1d2b1b2d1b, \\
c &\mapsto b2d1b2b1d2b1d2b1b2d1b, \\
d &\mapsto b2d1b2b1d2b1a2b1d2b1b2d1b, \\ 
1 &\mapsto 1, \qquad 2 \mapsto 2.
\end{align*}
Each of these is of the required form (that is, Type 1 edges alternate with
Type 2), so this is the required train-track. The corresponding transition
matrix (keeping only the finite-length Type 2 edges) for $\tau_1 \sigma_1
\rho_1^{-1} \sigma_1$ is then
\begin{equation}
\left[
\begin{array}{cccc}
0 & 0 & 0 & 1\\
18 & 10 & 7 & 8\\
2 & 1 & 0 & 0\\
7 & 4 & 4 & 4
\end{array}
\right]
\end{equation}
with spectral radius $14.48$, so the braid has a topological entropy of
$2.67$.  This growth rate is slightly greater than that achieved by performing 
the cylinder
braid~$\tau_1\sigma_1$ twice (resulting topological entropy~$2.63$), at the
same energetic cost (by symmetry).  However, no attempt has been made at
optimising this torus braid, so braids with higher topological entropy may
exist.

\section{Numerical Simulations}
\label{sec:numerics}

In this section we show numerical simulations of quasi-steady two-dimensional
spatially periodic Stokes flow to determine the sharpness of predicted entropy
lower bounds compared to actual flow topological entropies.

Before investigating a particular flow regime, it is important to remember
that the topological entropy bounds given above are based only on the braiding
motion of the stirrers and are independent of all properties of the underlying
fluid flow except for continuity.  In a given flow regime, however, we expect
that there could be additional features that lead to more intricate braiding,
and therefore to an improved topological entropy.
 
In our simulations, entropies are computed numerically by recording the
stretched length $L(t)$ of a material line as a function of time $t$. Because
material lines are stretched exponentially fast in a chaotic flow, the slope
of a best-fit line through $\log L(t)$ versus $t$ gives the topological
entropy of the flow (after ignoring initial transients).

We consider a very viscous (Stokes) fluid in a unit square domain $0 \le x,y <
1$. Stirring is performed with just two stirrers, since we have shown above
that this is sufficient to allow topological chaos in a periodic domain. The
stirrers are circular with diameter $1/10$, and are initially centred at
$(\tfrac{1}{4},\tfrac{1}{4})$ and $(\tfrac{3}{4},\tfrac{3}{4})$. In the
stirring operations $\sigma_1$, $\tau_1$ and $\rho_1$ the stirrer trajectories
we prescribe are given in Table~\ref{tab:rodtraj}.  Note that in the
Stokes flow regime the resultant advection depends only upon the paths of the
stirrers and not on the stirrer velocities.
\begin{table}
\caption{Trajectories of the stirrers in the simulations corresponding to the
  braid generators, with~\hbox{$0\le\alpha<1$} and~\hbox{$\theta =
  \left(\tfrac{1}{4}-\alpha\right) \pi$}.}
\begin{tabular}{lcccc}
\\
\hline\hline
generator && stirrer 1 && stirrer 2\\
\hline
$\sigma_1$ & \qquad\qquad\qquad & 
$\left(\tfrac{1}{2}-\tfrac{1}{2\sqrt{2}} \cos\theta,
\tfrac{1}{2}-\tfrac{1}{2\sqrt{2}}\sin\theta\right)$ & \qquad\qquad &
$\left(\tfrac{1}{2}+\tfrac{1}{2\sqrt{2}} \cos\theta,
\tfrac{1}{2}+\tfrac{1}{2\sqrt{2}} \sin\theta\right)$ \\[4pt]
$\tau_1$ && $\left(\tfrac{1}{4}-\alpha,\tfrac{1}{4}\right) \mod 1$ &&
$\left(\tfrac{3}{4},\tfrac{3}{4}\right)$ \\[4pt]
$\rho_1$ && $\left(\tfrac{1}{4},\tfrac{1}{4}-\alpha\right) \mod 1$ &&
$\left(\tfrac{3}{4},\tfrac{3}{4} \right)$\\[4pt]
\hline\hline
\end{tabular}
\label{tab:rodtraj}
\end{table}

The Stokes flow velocity field was computed using the lattice-Boltzmann
method~\cite{Chen1998}. Whilst a lattice-Boltzmann code is relatively
inefficient for computing Stokes flow, it copes easily with the moving
circular boundaries and also with periodic boundary conditions, in addition to
allowing investigation of similar flows at moderate Reynolds number (not
reported here). We have chosen to compute the velocity field this way because
the complex variable series solution approach employed by
Finn~\etal~\cite{Finn2003a} does not generalise easily for periodic
domains. In computing the velocity field, for simplicity we impose a spatially
periodic pressure field, and thus find the unique solution that has no net
flux in either of the periodic directions.

We consider passive advection of an initial closed polygon with corners at
$(0,0)$, $(\tsfrac12,0)$, $(\tsfrac12,1)$, $(1,1)$, $(1,\tsfrac12)$ and
$(0,\tsfrac12)$, with a total length of $L(0)=4$. This represents the
interface between a two-by-two checkerboard pattern of black and white squares
of fluid.  As the interface is advected we use a particle insertion scheme to
ensure that stretches and folds remain well resolved~\cite{Krasnopolskaya1999}.

\begin{figure}
\begin{center}
\parbox[t]{3cm}{\epsfig{file=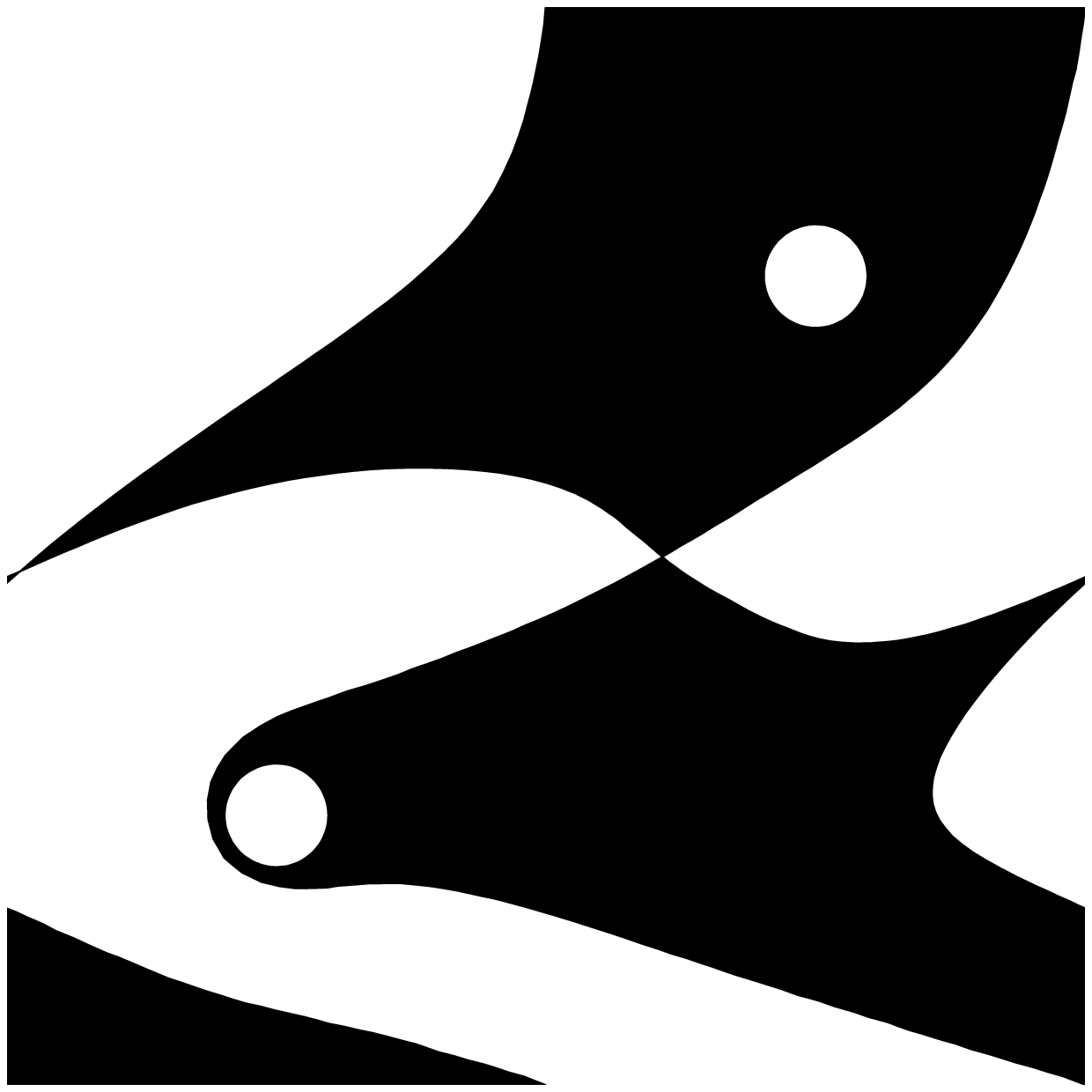,height=3cm}\\ \leftline{$t=1/2$}}
\parbox[t]{3cm}{\epsfig{file=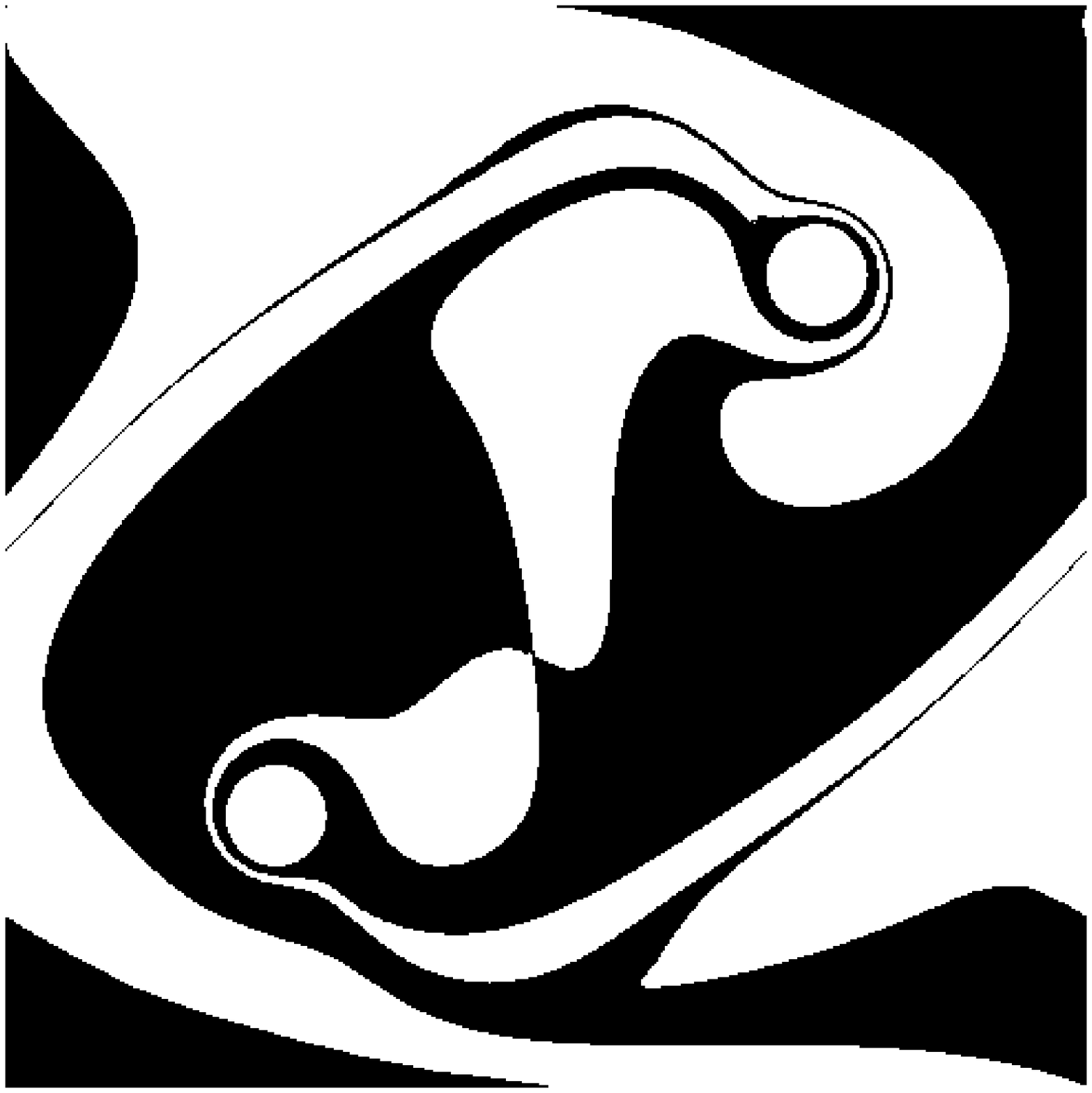,height=3cm}\\ \leftline{$t=1$}}
\parbox[t]{3cm}{\epsfig{file=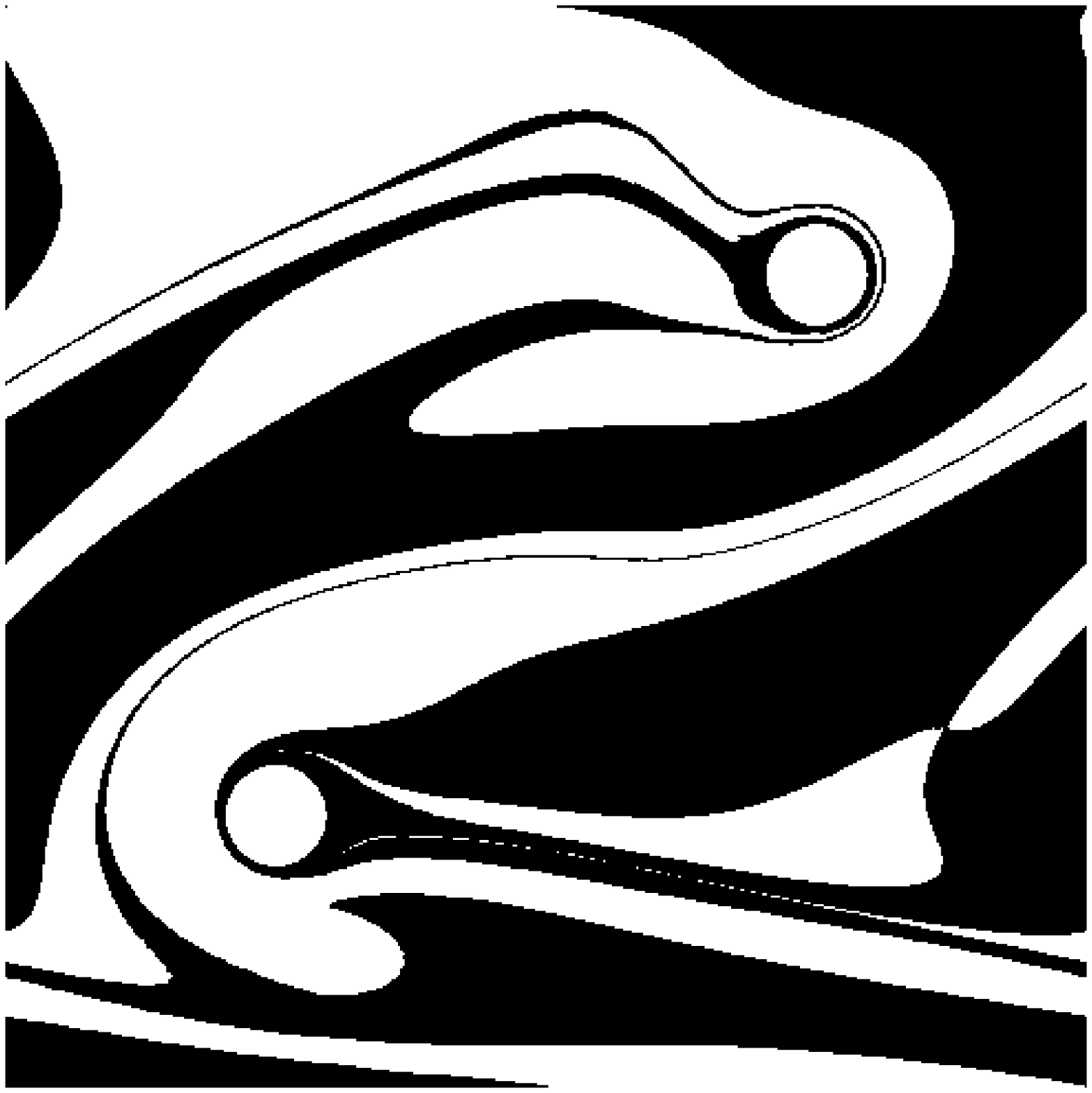,height=3cm}\\ \leftline{$t=3/2$}}
\parbox[t]{3cm}{\epsfig{file=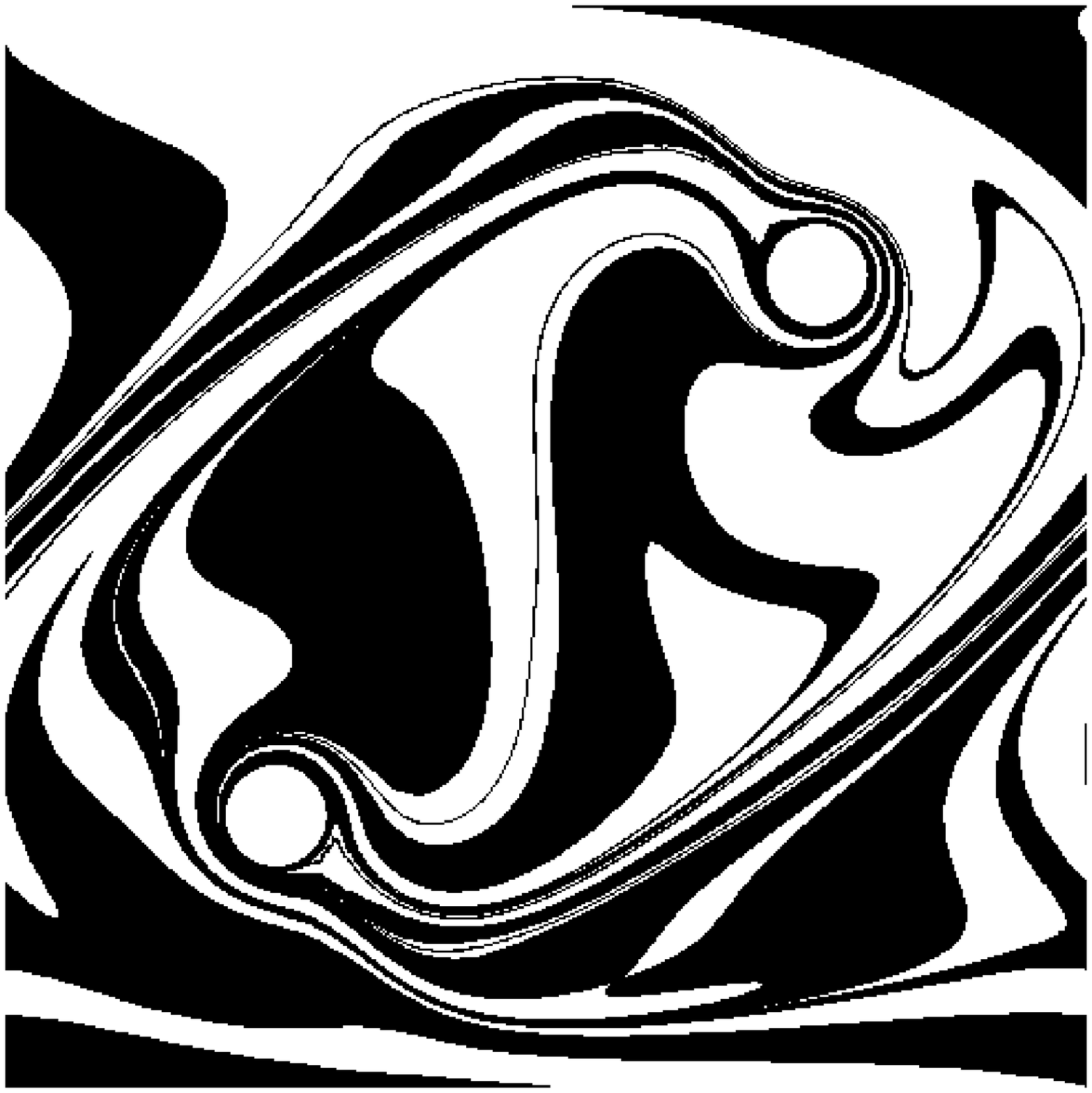,height=3cm}\\ \leftline{$t=2$}}

\smallskip

\parbox[t]{3cm}{\epsfig{file=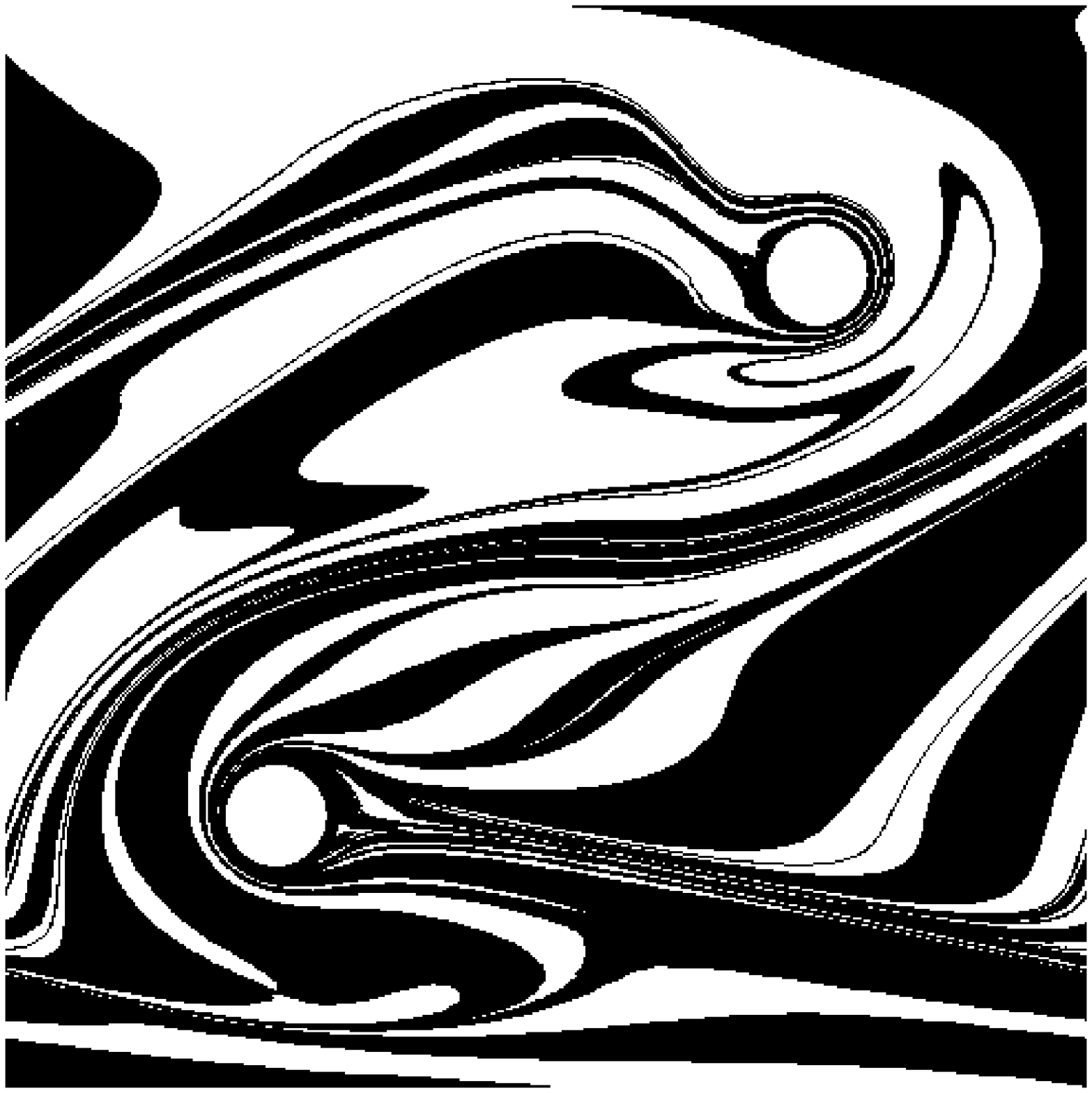,height=3cm}\\ \leftline{$t=5/2$}}
\parbox[t]{3cm}{\epsfig{file=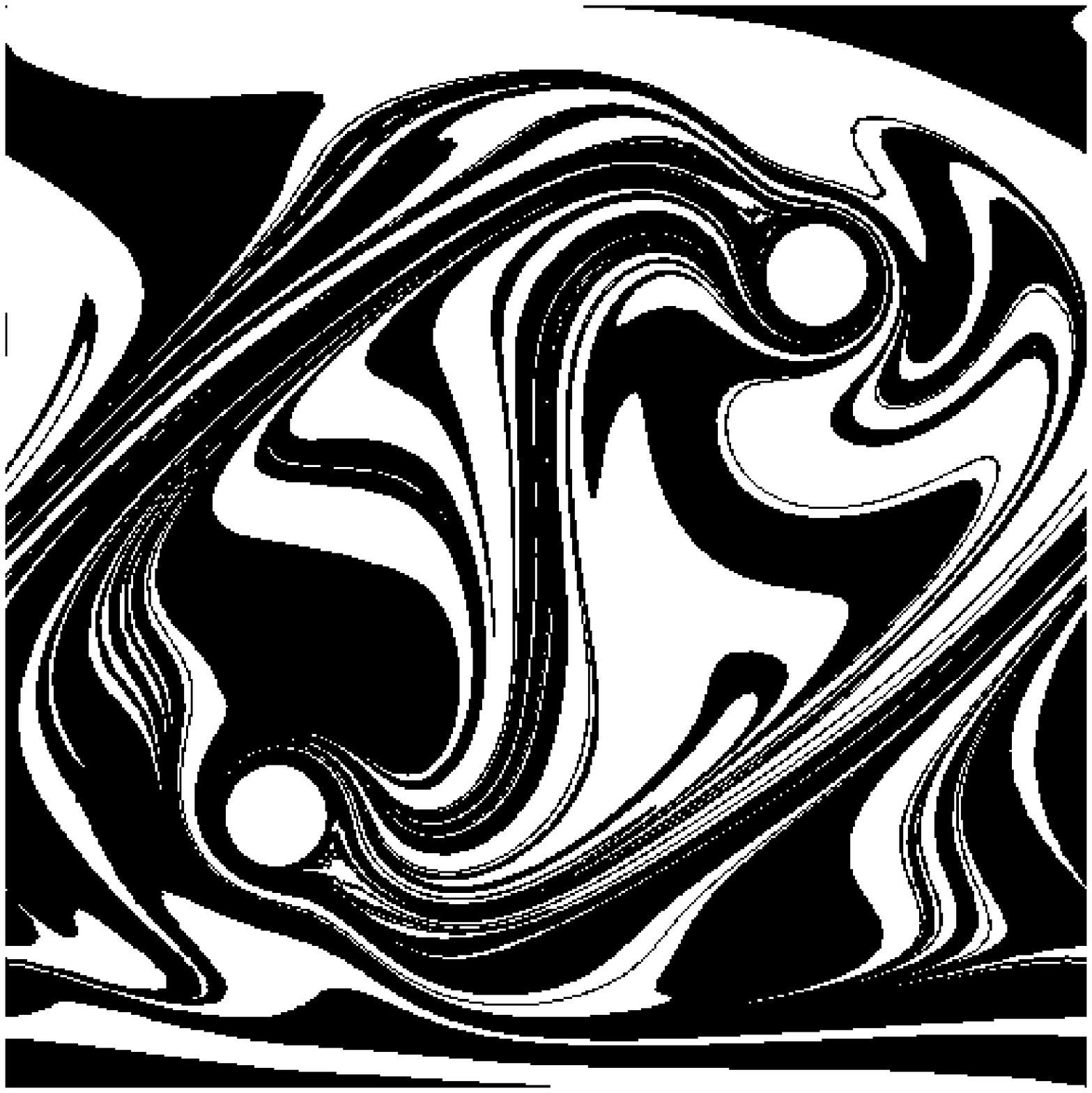,height=3cm}\\ \leftline{$t=3$}}
\parbox[t]{3cm}{\epsfig{file=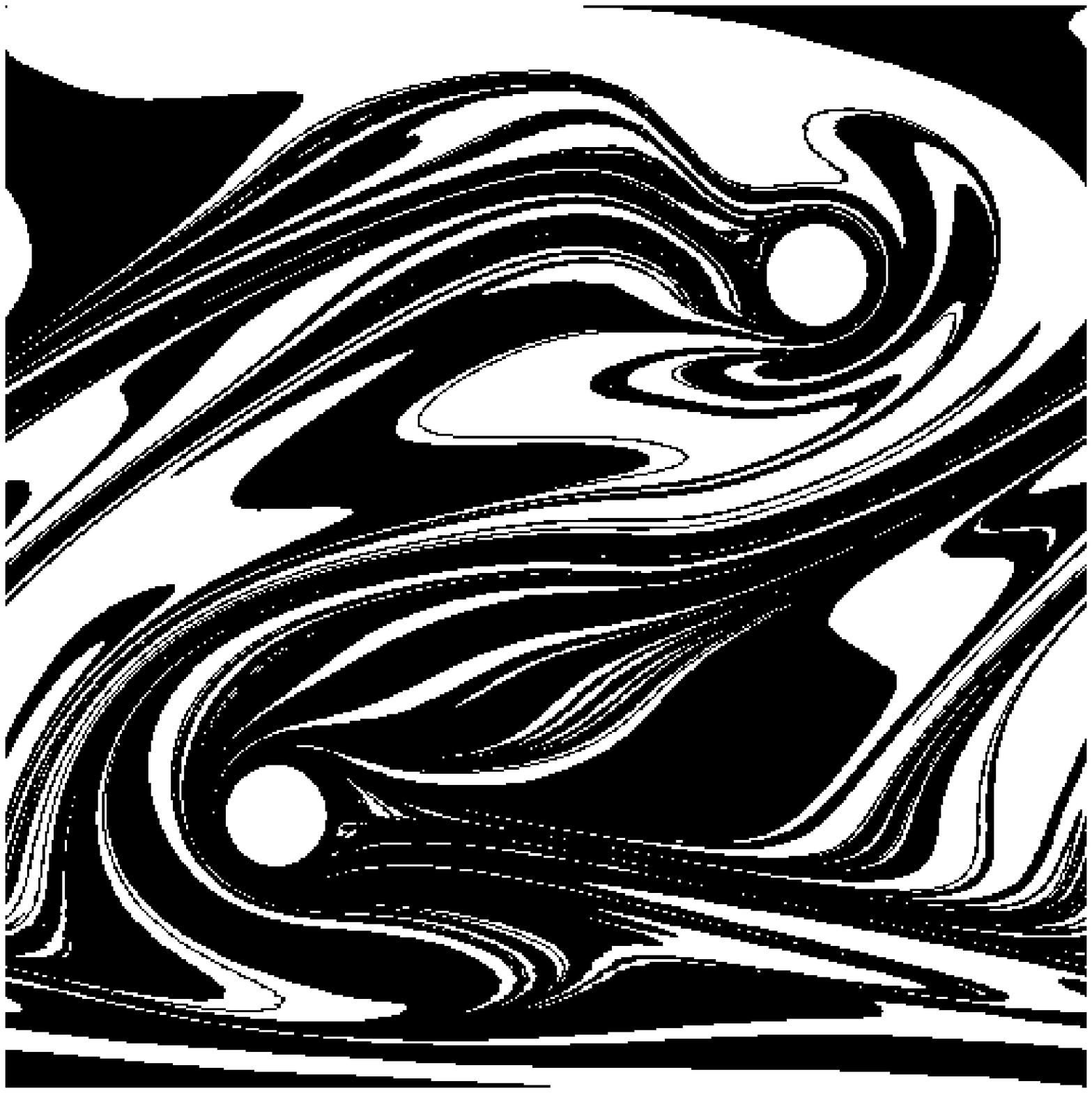,height=3cm}\\ \leftline{$t=7/2$}}
\parbox[t]{3cm}{\epsfig{file=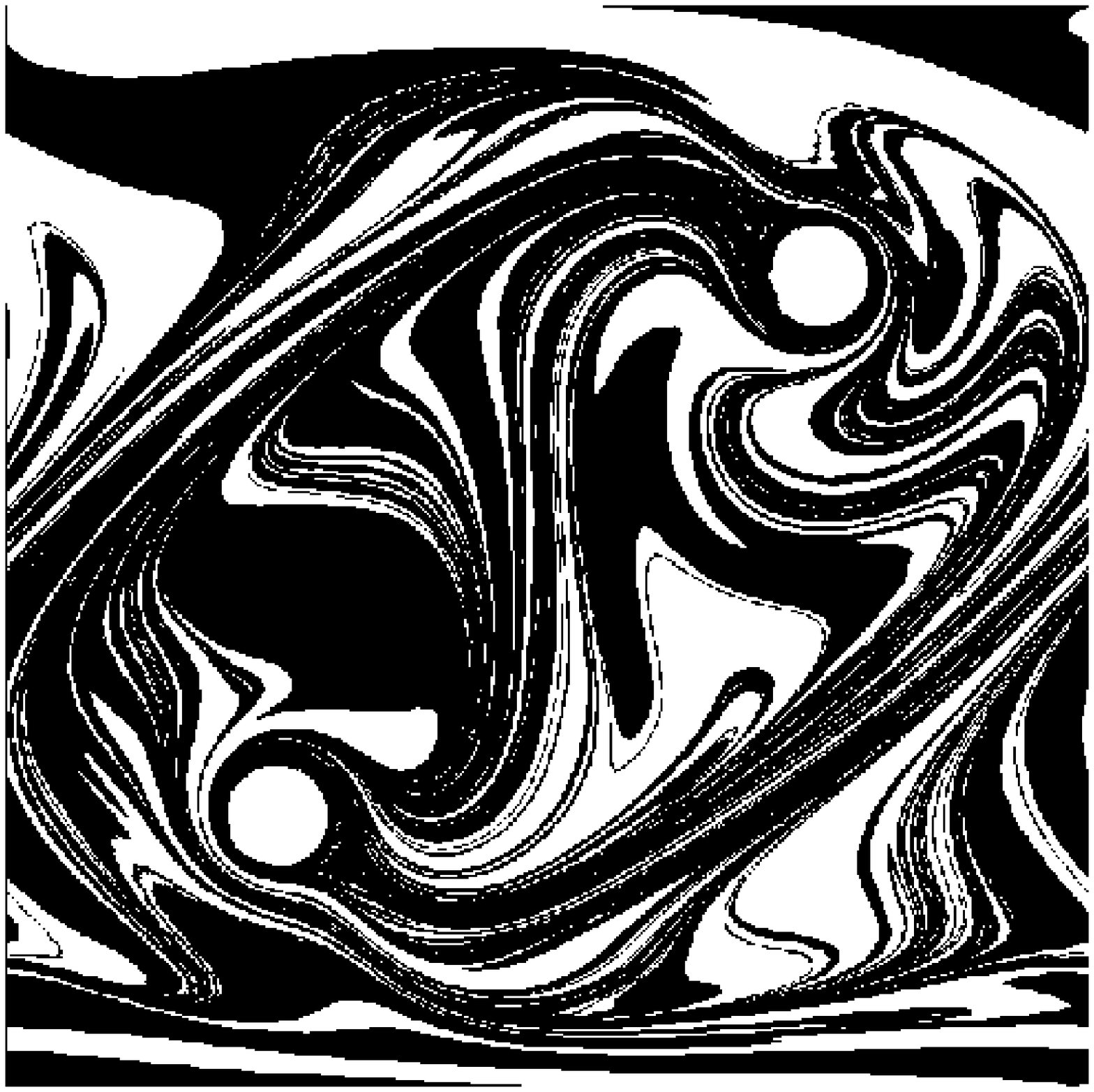,height=3cm}\\ \leftline{$t=4$}}
\end{center}
\caption{Time-periodic stirred flow on the cylinder under the braid $\tau_1
\sigma_1$. Each snapshot represents a half-period.}
\label{fig:singly}
\end{figure}

\begin{figure}
\begin{center}
\parbox[t]{3cm}{\epsfig{file=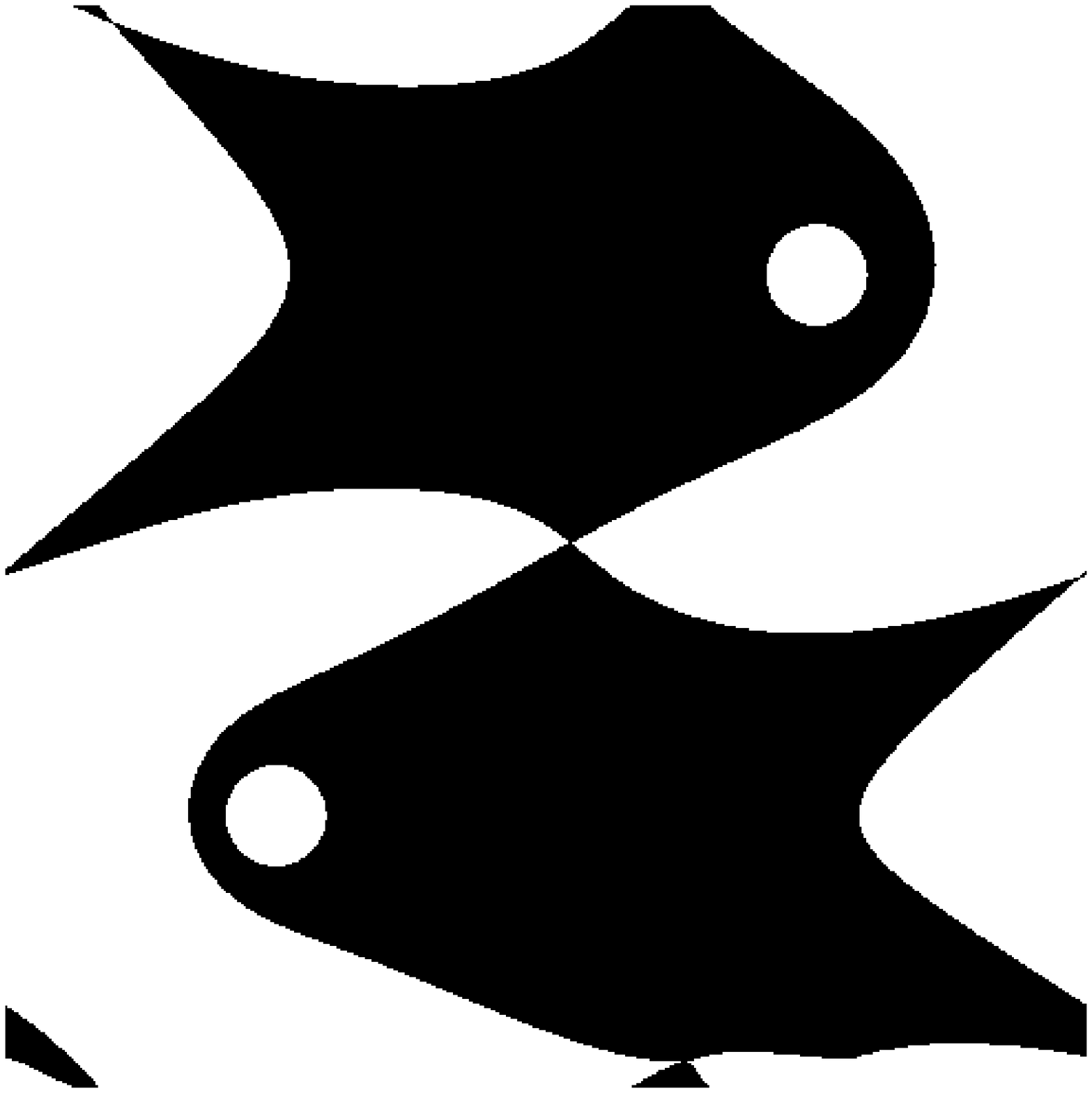,height=3cm}\\ \leftline{$t=1/4$}}
\parbox[t]{3cm}{\epsfig{file=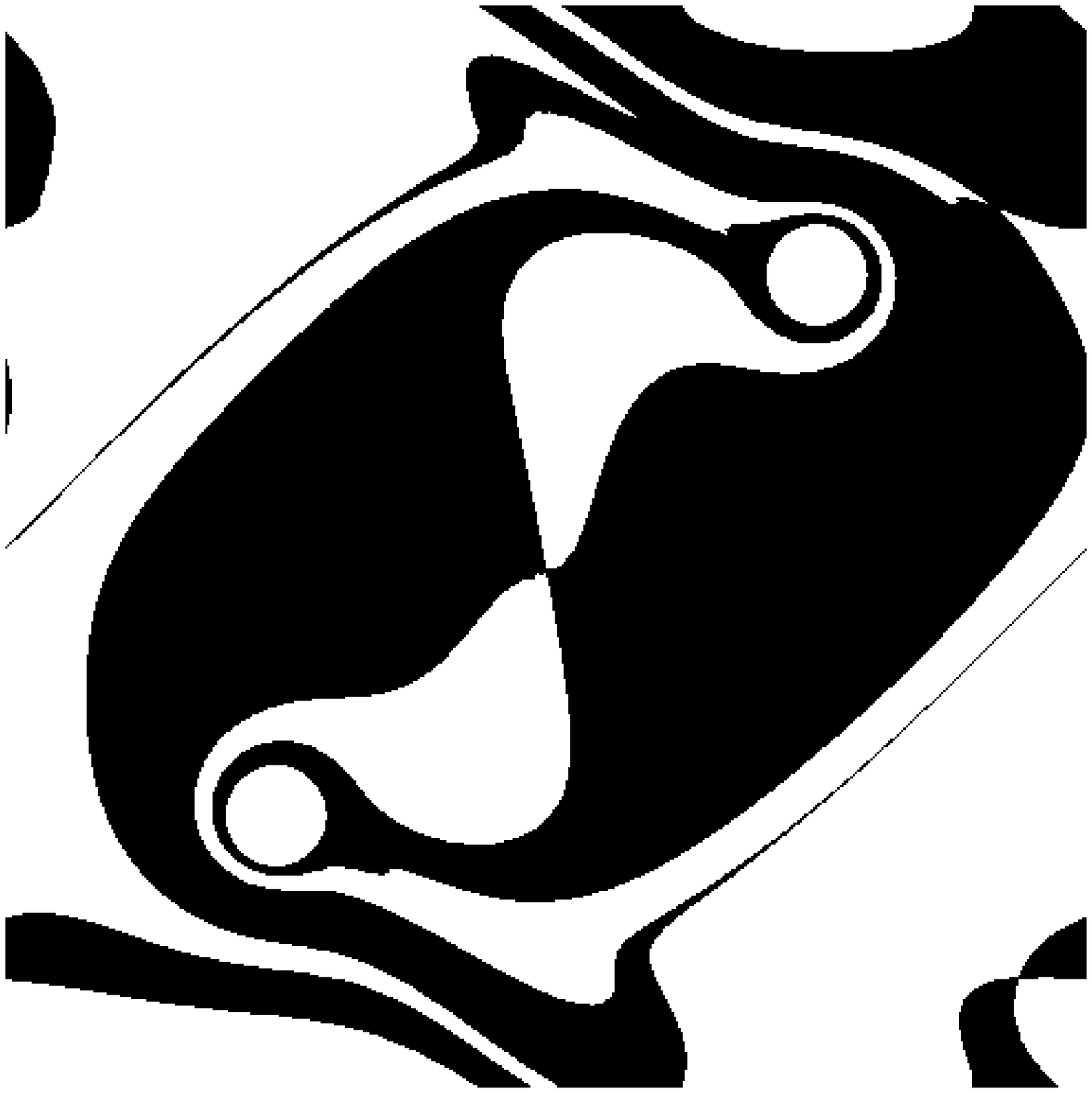,height=3cm}\\ \leftline{$t=2/4$}}
\parbox[t]{3cm}{\epsfig{file=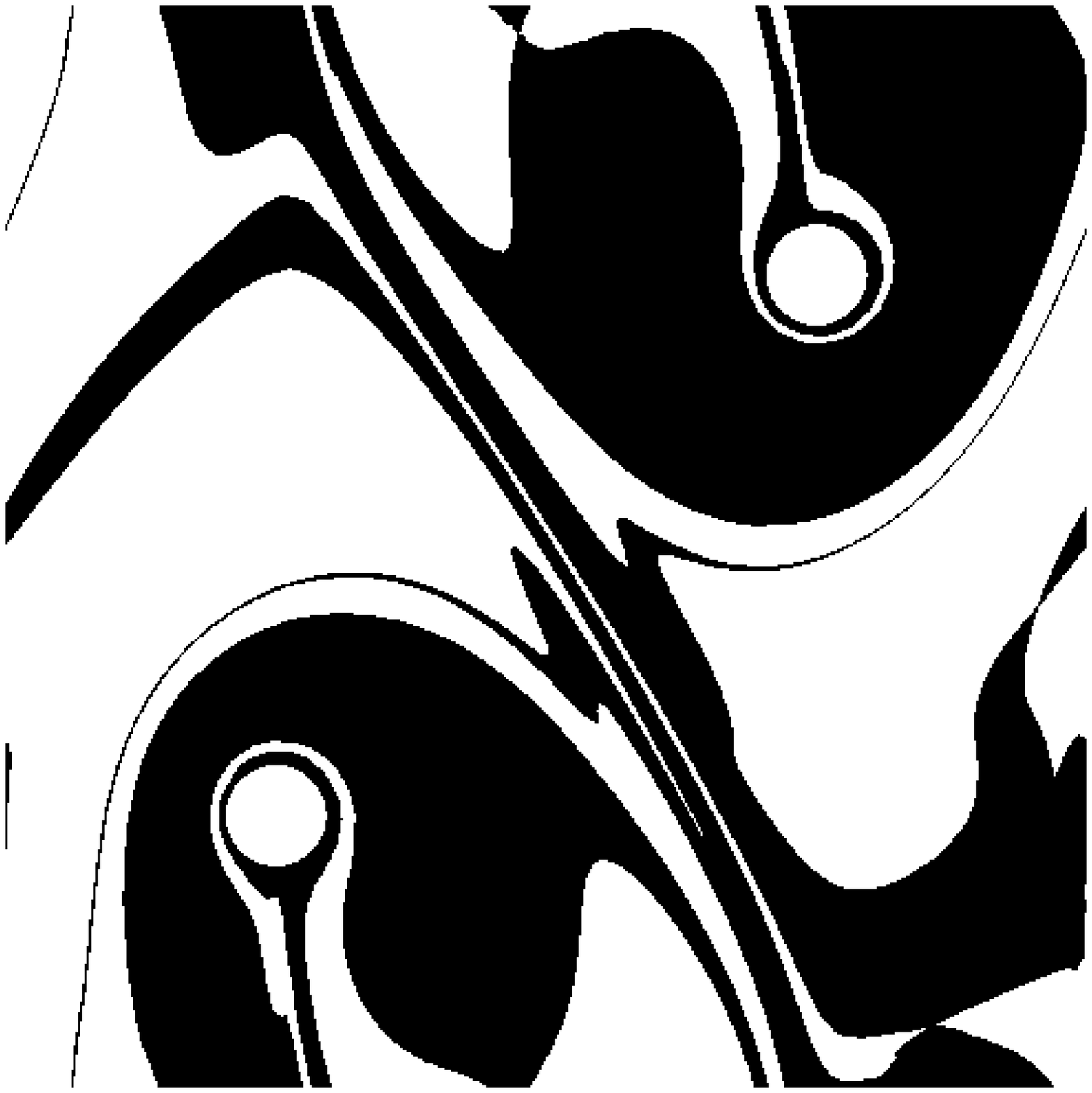,height=3cm}\\ \leftline{$t=3/4$}}
\parbox[t]{3cm}{\epsfig{file=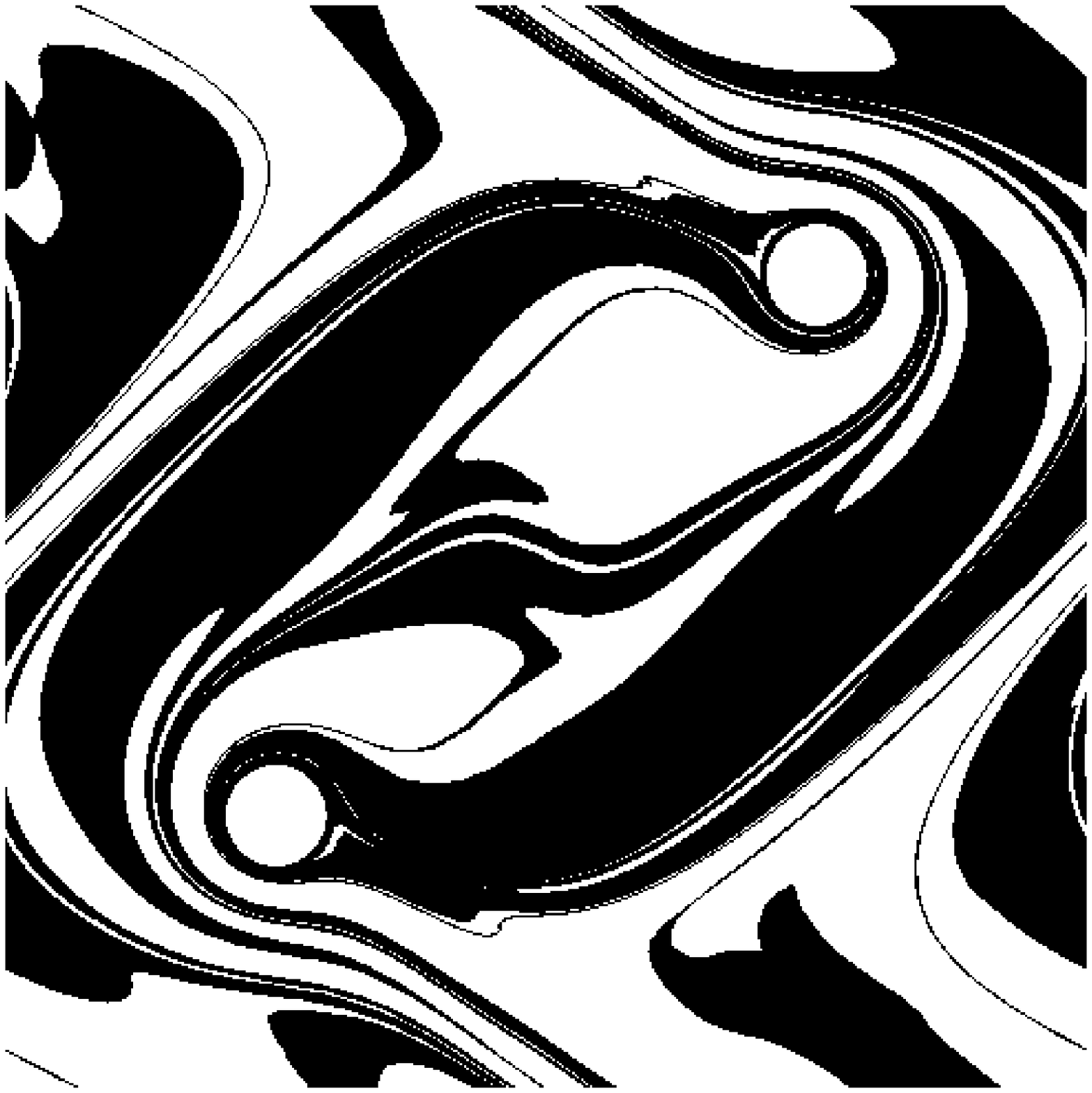,height=3cm}\\ \leftline{$t=1$}}

\smallskip

\parbox[t]{3cm}{\epsfig{file=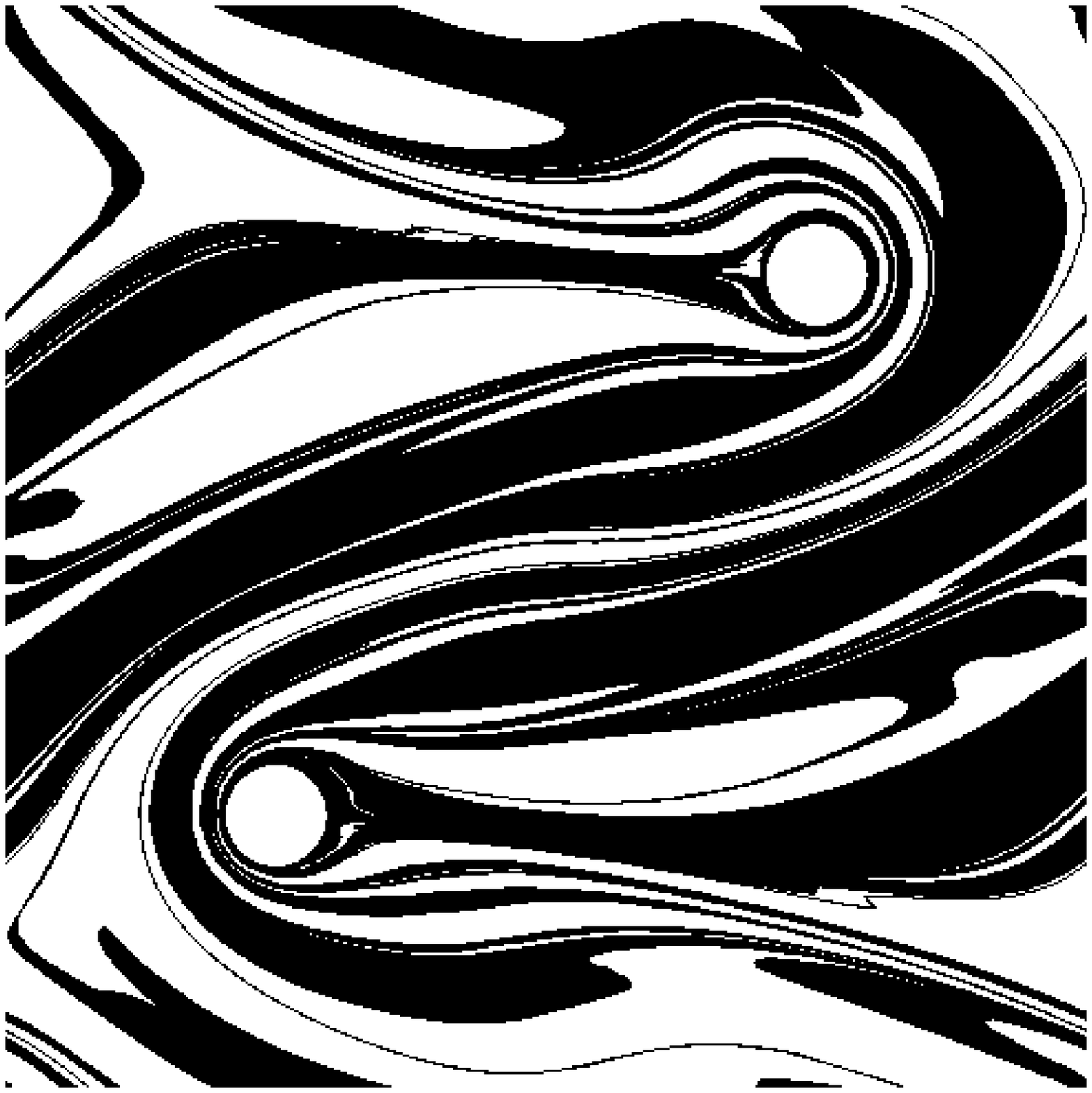,height=3cm}\\ \leftline{$t=5/4$}}
\parbox[t]{3cm}{\epsfig{file=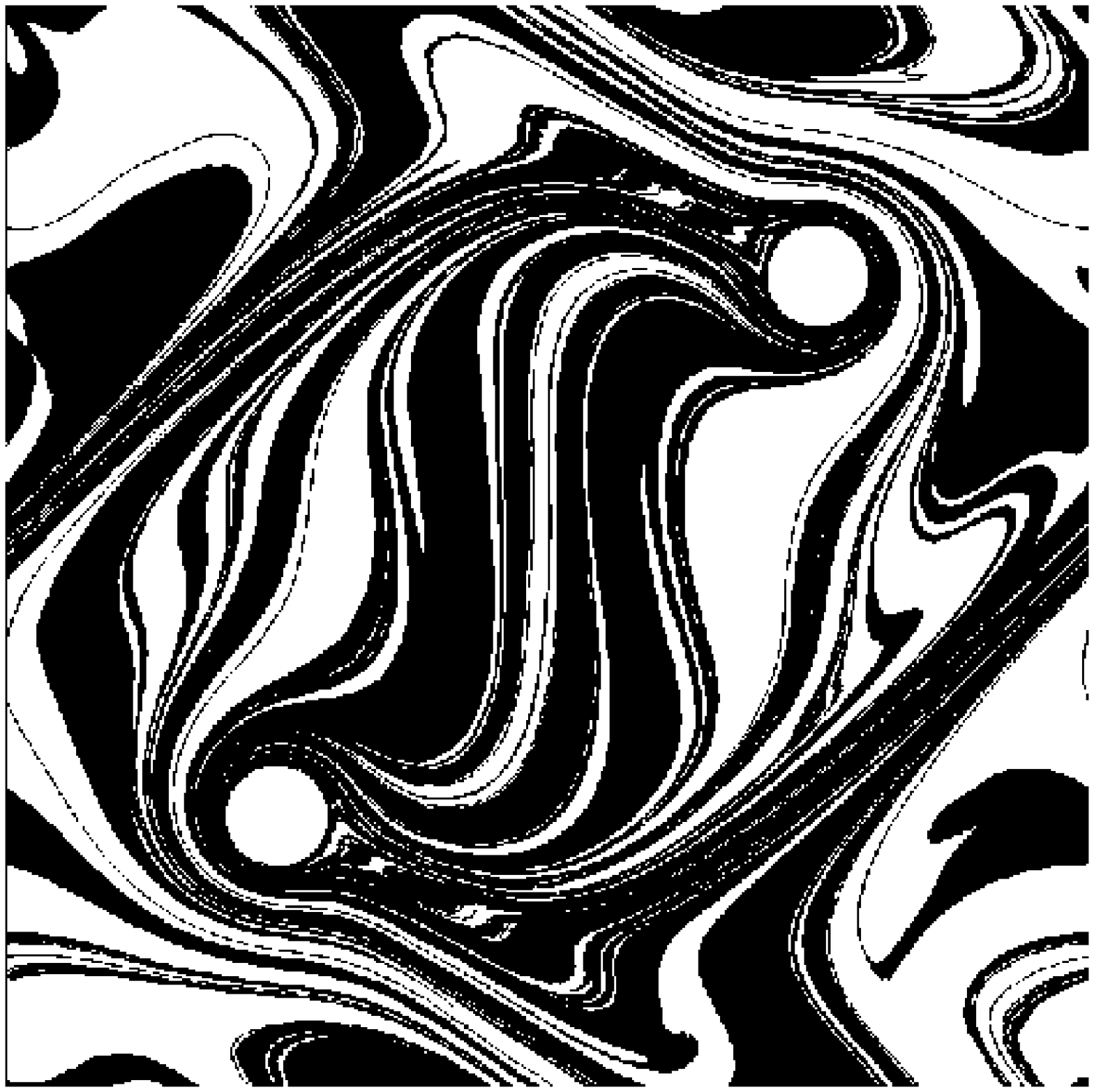,height=3cm}\\ \leftline{$t=6/4$}}
\parbox[t]{3cm}{\epsfig{file=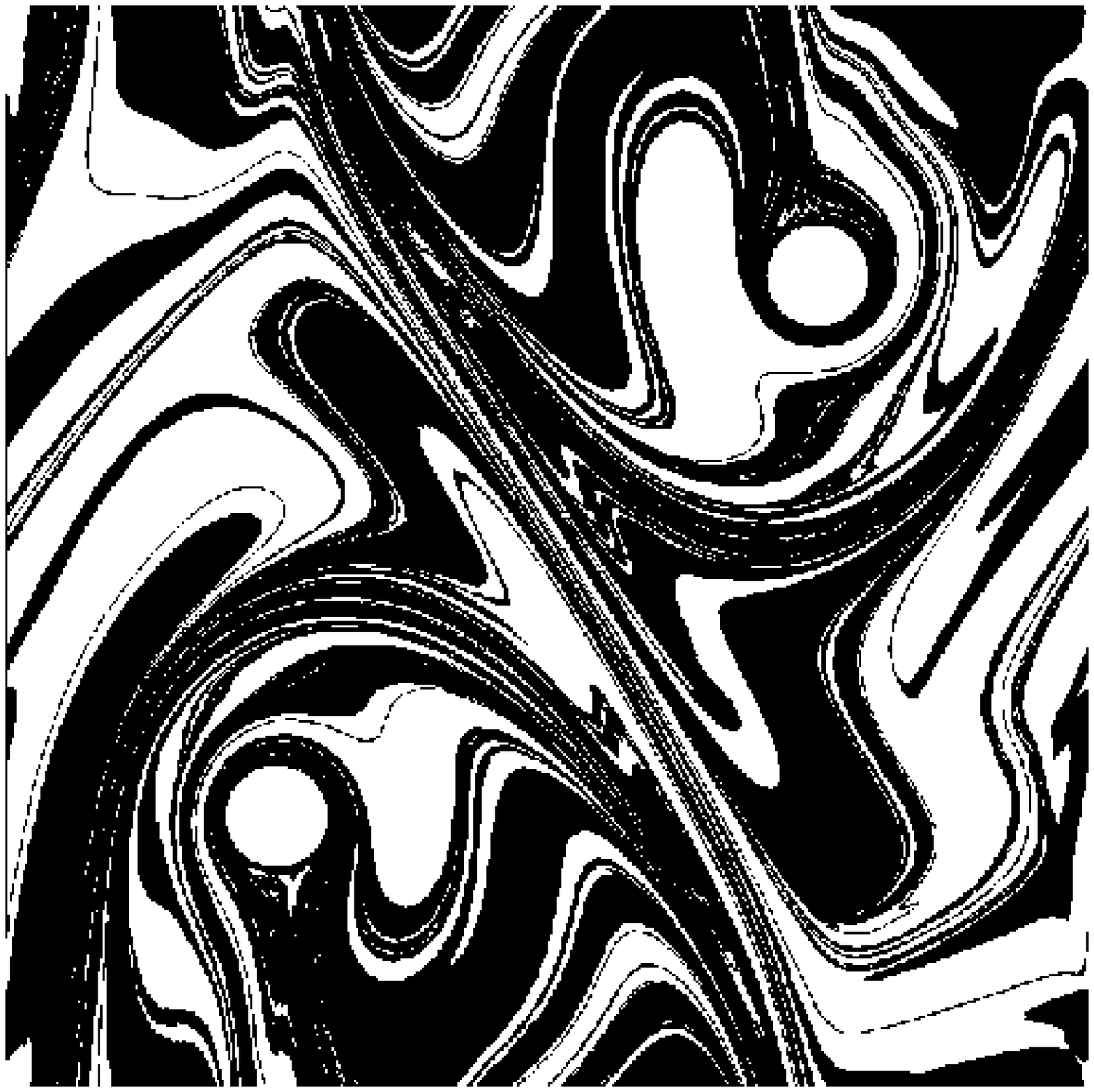,height=3cm}\\ \leftline{$t=7/4$}}
\parbox[t]{3cm}{\epsfig{file=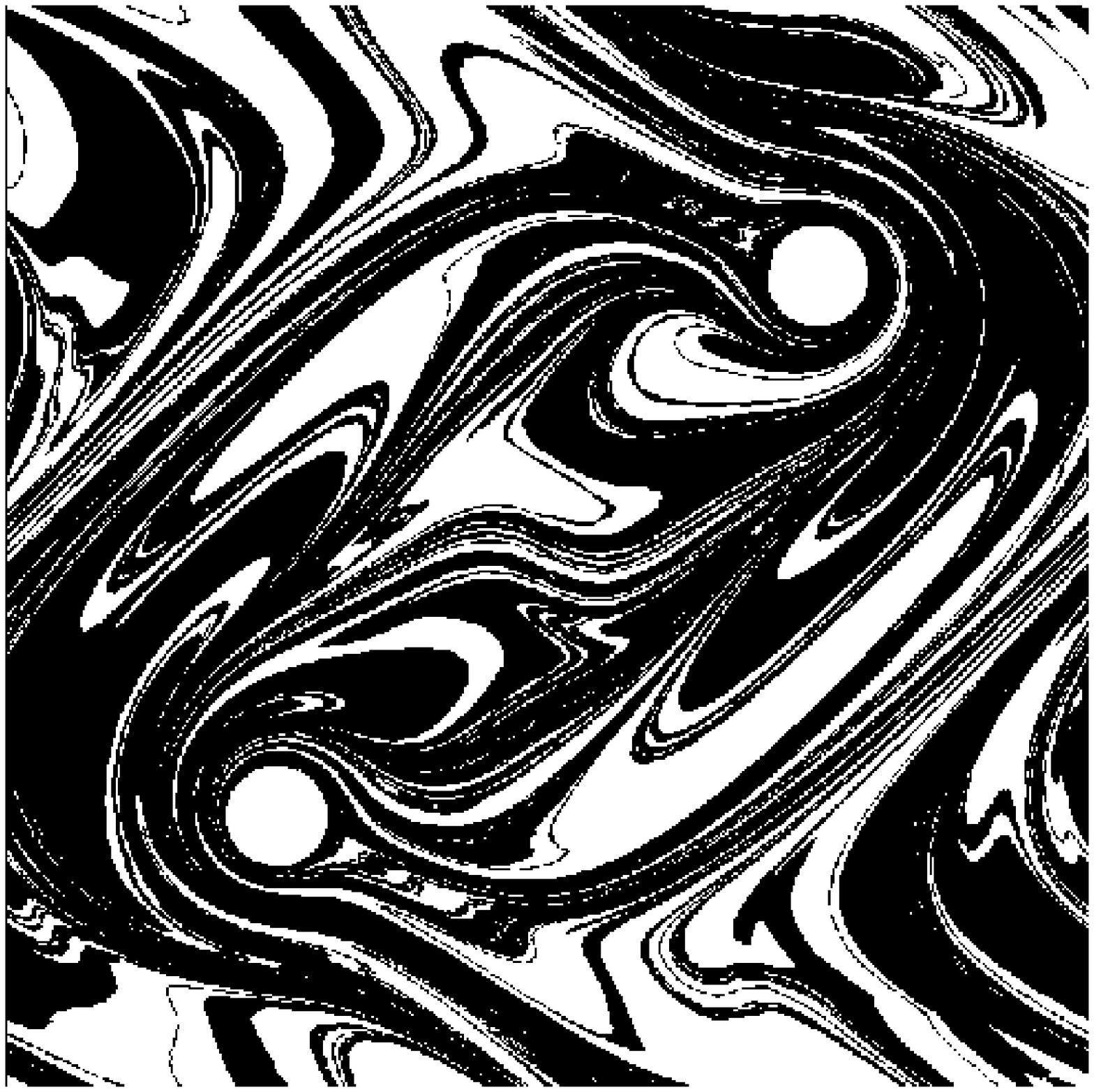,height=3cm}\\ \leftline{$t=2$}}
\end{center}
\caption{Time-periodic stirred flow on the torus under the braid 
$\tau_1 \sigma_1 \rho_1^{-1} \sigma_1$. Each snapshot represents
a quarter-period.}
\label{fig:doubly}
\end{figure}

In Figure~\ref{fig:singly} we show the result of stirring using the braid
$\tau_1\sigma_1$ in a domain that is periodic in the $x$-direction, with
no-slip boundaries at $y=0,1$. Figure~\ref{fig:doubly} shows the evolution of
the fluid under the braid $\tau_1 \sigma_1 \rho_1^{-1} \sigma_1$ in a
doubly-periodic domain.

In both figures it is clear that the fluid is well mixed. Closer inspection of
the snapshots after complete application of each braid also reveals
accumulations of thin striations that roughly align with the train-track
graph. This is more obvious for the singly-periodic case shown in
Figure~\ref{fig:singly}, where the thinnest striations form in a `Z' shape
similar to that shown in the corresponding train-track in
Figure~\ref{fig:strack}. It was noted by Alvarez {\it et
al.}~\cite{Alvarez1998} that in chaotic flows striations accumulate with
different densities in different regions of the domain.  Here, in the
long-time limit the relative number of striations caught between stirrers is
given by the entries of the dominant eigenvector of the train-track transition
matrix.  This relative number is slightly different from the density, which
depends on the size of the domain and could be locally nonuniform, but it
should correlate well with it.  We have not attempted to verify this, since it
would require a separate study.

\begin{figure}
\begin{center}
\psfrag{log L\(t\) / L\(0\)}{\hspace{-1em}\raisebox{.5em}{$\log [L(t)/L(0)]$}}
\psfrag{t/T}{\raisebox{-.4em}{$t/T$}}
\psfrag{torus}{$\!\!\!\tau_1\sigma_1\rho_1^{-1}\sigma_1$ 2.7}
\psfrag{cylinder}{$\tau_1\sigma_1$ 1.4}
\psfrag{sine}{sine 1.4}
\epsfig{file=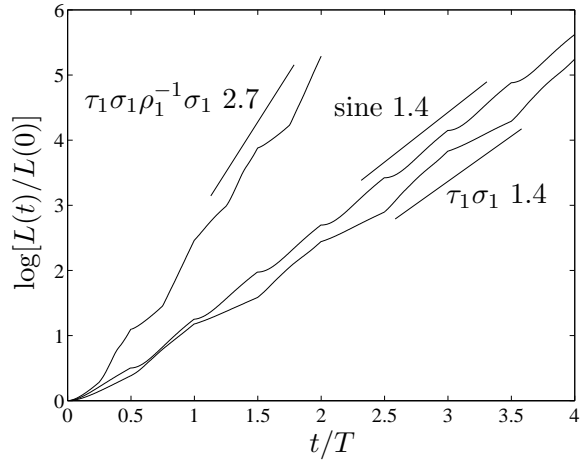,height=6cm}
\end{center}
\caption{Interface length for all the model flows. Plot of interface length
  versus time for braids in Figures \ref{fig:singly} and \ref{fig:doubly}. The
  length is shown on a log scale. The slope of each curve for large time gives
  the topological entropy of each flow.}
\label{fig:lengths}
\end{figure}

To quantify the mixing in our two model flows we calculate numerically their
topological entropies. Figure~\ref{fig:lengths} shows the interface length
$L(t)$ for the two flows in Figures~\ref{fig:singly} and \ref{fig:doubly}.
The curves quickly reach an exponential growth regime, though there is clearly
some variation in growth rate during each period of the flow. To allow for
this we calculate the slope of the curves from a linear least-squares fit,
ignoring transient data in the first period of the flow. There is clearly some
subjectivity about how such a fit is performed, so we claim our slopes are
accurate to within ten percent.  We find that the $\tau_1\sigma_1$ flow has an
entropy of $1.4\pm10\%$, so the lower bound of $1.32$ derived in
Section~\ref{sec:periodic}, based only on the braiding of the stirrers, is
around $90\%$ sharp. For the doubly-periodic braid $\tau_1 \sigma_1
\rho_1^{-1} \sigma_1$ our numerically computed entropy is around $2.7\pm10\%$,
meaning that the lower-bound of $2.67$ is at least $90\%$ sharp. The sharpness
of the bounds indicates that the braiding motion of the stirrers in our two
model flows are responsible for almost all of the chaos, and only a small
amount of chaos is due to secondary effects such as periodic orbits.

\section{The Sine Flow}
\label{sec:sineflow}

The fact that fluid particles themselves act as topological obstacles to flow
in two dimensions~\cite{Thiffeault2005} means that rigorous topological
entropy lower bounds can be computed even when stirring apparatus is not
present to force a braiding motion. To provide an illustration we consider the
well-known sine flow~\cite{Pierrehumbert1994}, which has not been
examined from such a topological perspective.

The doubly-periodic sine flow is defined on the unit square $0 \le x,y <
1$. We consider, for simplicity, the case where the velocity field has a
temporal period of unity, with the velocity field given by $(\sin 2\pi y,0)$
for $0\le t<\tsfrac12$ and $(0,\sin 2\pi x)$ for $\tsfrac12 \le t< 1$. It is
easy to verify that in this case the trajectories of the initial points
$(\tsfrac14,0)$, $(\tsfrac34,0)$, $(\tsfrac14,\tsfrac12)$,
$(\tsfrac34,\tsfrac12)$, $(0,\tsfrac14)$, $(0,\tsfrac34)$,
$(\tsfrac12,\tsfrac14)$ and $(\tsfrac12,\tsfrac34)$ are all periodic, with
period $2$.

\begin{figure}
\begin{center}
\parbox[t]{3cm}{\epsfig{file=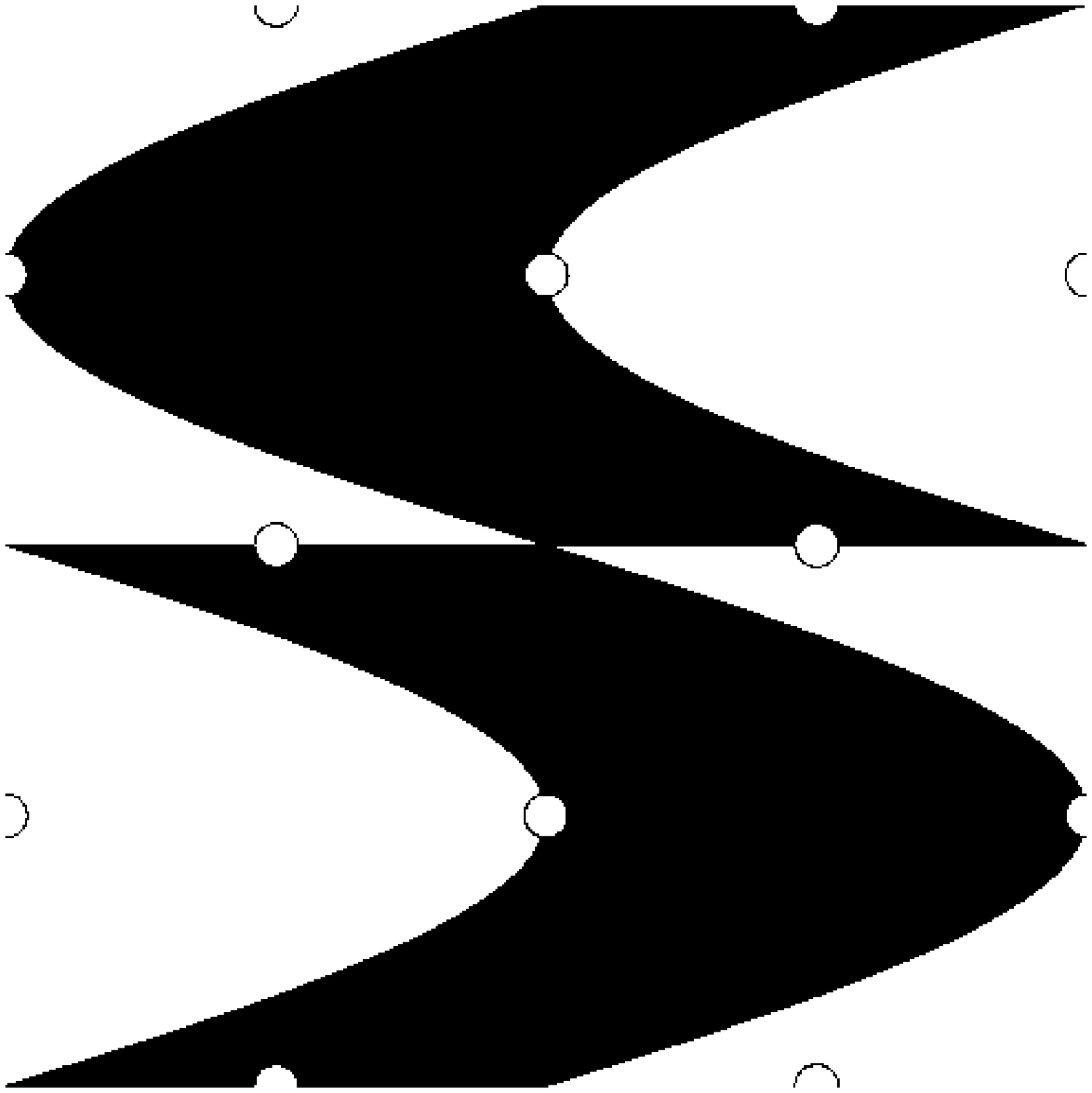,height=3cm}\\ \leftline{$t=1/2$}}
\parbox[t]{3cm}{\epsfig{file=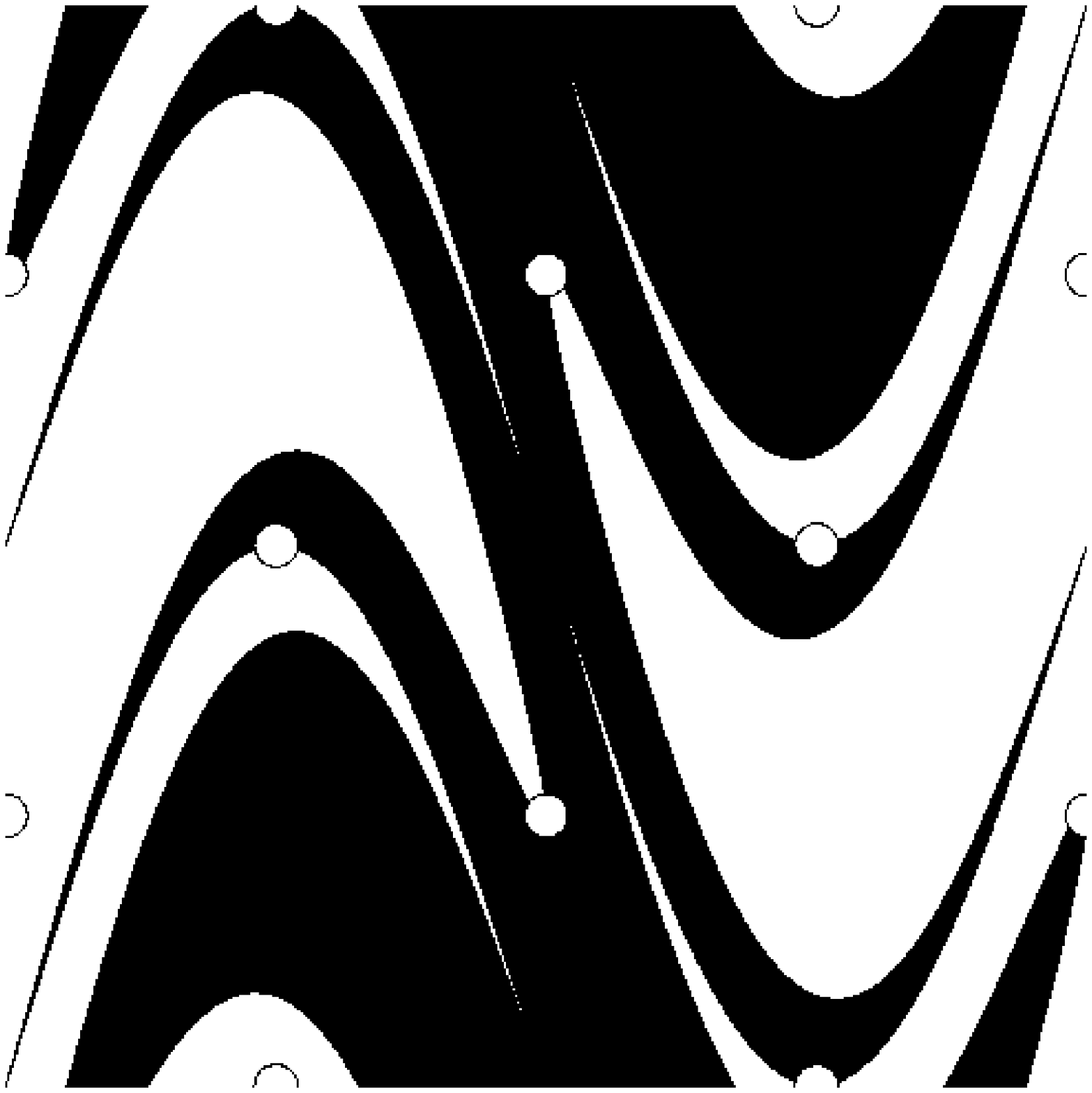,height=3cm}\\ \leftline{$t=1$}}
\parbox[t]{3cm}{\epsfig{file=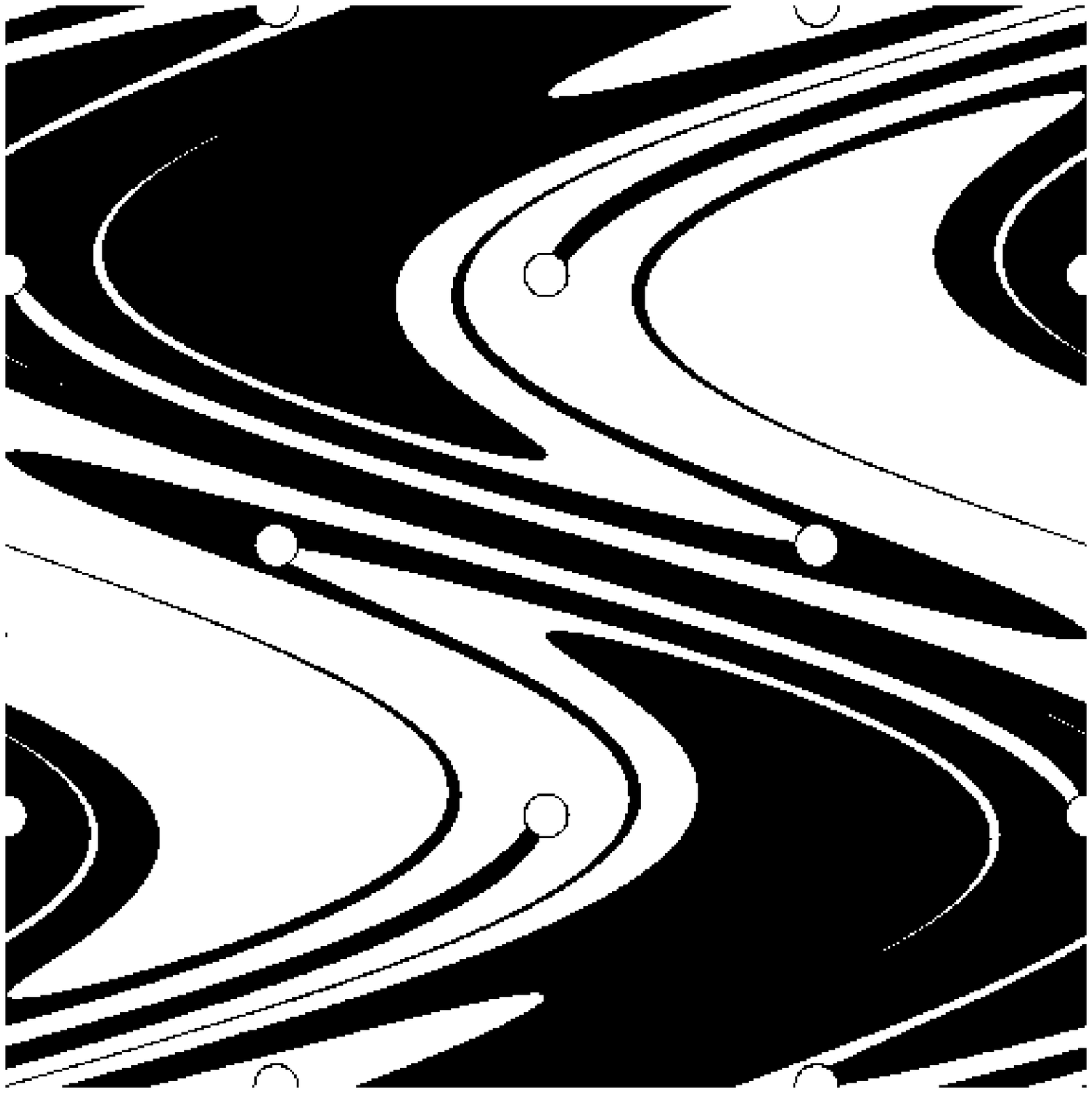,height=3cm}\\ \leftline{$t=3/2$}}
\parbox[t]{3cm}{\epsfig{file=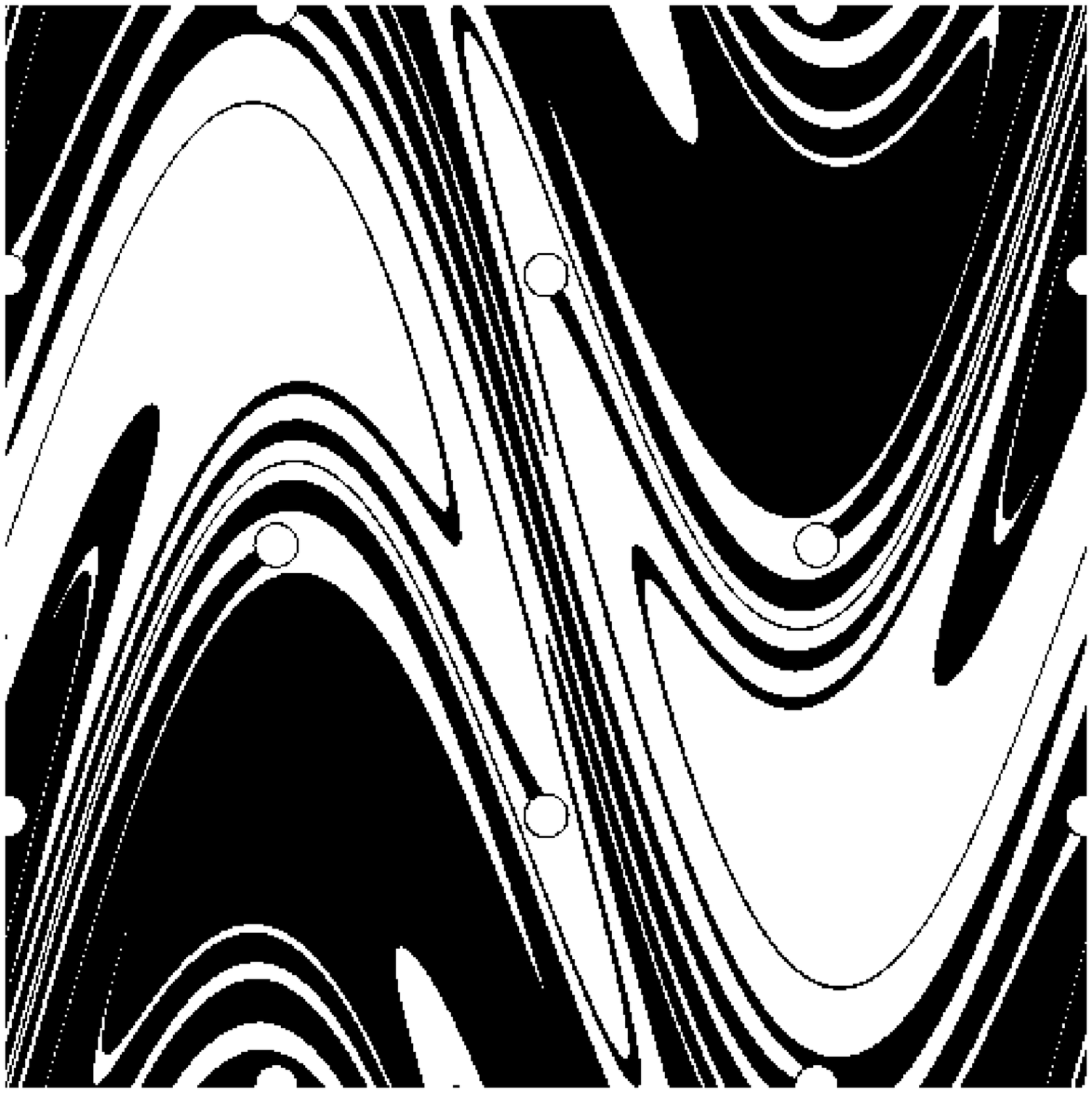,height=3cm}\\ \leftline{$t=2$}}
\end{center}
\caption{Sine flow advection simulations. Each snapshot represents a
half-period.  The circles denote periodic orbits, with period 2.}
\label{fig:sineexamples}
\end{figure}

Figure~\ref{fig:sineexamples} shows the result of advecting the initial
condition described in Section~\ref{sec:numerics} according to the sine flow.
It is easy to see from this numerical simulation that the trajectories of
these points form a non-trivial braid. One way of writing this braid is
$\sigma_1 \sigma_2^{-1} \tau_4^{-1} \sigma_3^{-1} \sigma_2^{-1} \sigma_1^{-1}
\sigma_7 \sigma_6^{-1} \tau_5 \sigma_5^{-1} \sigma_6^{-1} \sigma_7^{-1}
\rho_3^{-1} \sigma_2 \rho_6 \sigma_6$.
Note how highly-curved folds of the interface form around some of the periodic
points, showing that these points really do act like small stirrers! This can
occur because these are parabolic points with a singular unstable manifold,
allowing material lines to fold around the point.  The nature of these points
and the folding behaviour will be addressed in future work.

\begin{figure}
\begin{center}
\psfrag{edge1}{$1$}
\psfrag{edge2}{$2$}
\psfrag{edge3}{$4$} 
\psfrag{edge4}{$3$}
\psfrag{edgea}{$a$}
\psfrag{edgeb}{$b$}
\psfrag{edgec}{$c$}
\psfrag{edged}{$d$}
\psfrag{edgee}{$e$}
\psfrag{edgef}{$f$}
\epsfig{file=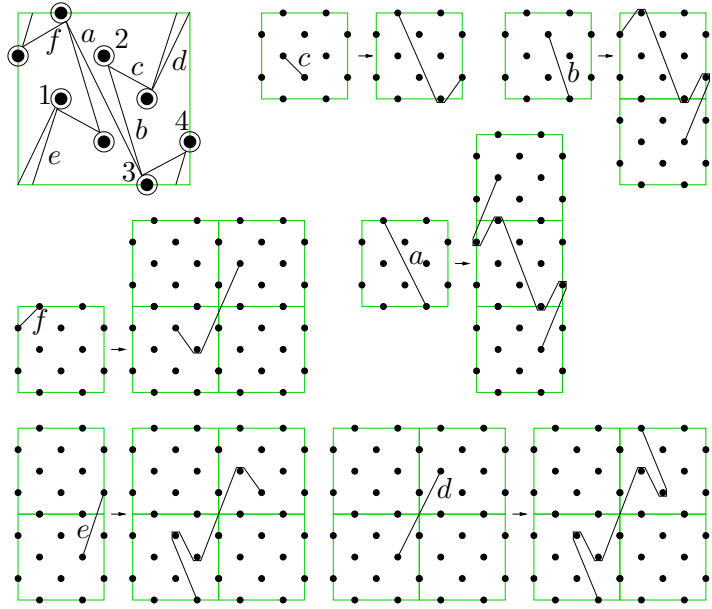,scale=0.3}
\end{center}
\caption{Train-track for the sine Flow.}
\label{fig:sinetrack}
\end{figure}

In Figure~\ref{fig:sinetrack} we show the construction of the train-track for
this braid, calculated in the same way as described in
Section~\ref{sec:periodic}.  The snapshots of material lines in
Figure~\ref{fig:sineexamples} are again invaluable in determining the
train-track.  The edge to edge-path mapping is
\begin{align*}
a &\mapsto e4f3a3f4e, \\
b &\mapsto f3a3f4e, \\
c &\mapsto a3f, \\
d &\mapsto b2c1d1c2b, \\
e &\mapsto b2c1d1c, \\
f &\mapsto c1d, \\
1 &\mapsto 3, \quad
2 \mapsto 4, \quad
3 \mapsto 1, \quad
4 \mapsto 2,
\end{align*}
with the corresponding transition matrix
\begin{equation}
\left[
\begin{array}{cccccc}
1 & 1 & 1 & 0 & 0 & 0 \\
0 & 0 & 0 & 2 & 1 & 0 \\
0 & 0 & 0 & 2 & 2 & 1 \\
0 & 0 & 0 & 1 & 1 & 1 \\
2 & 1 & 0 & 0 & 0 & 0 \\
2 & 2 & 1 & 0 & 0 & 0
\end{array}
\right]
\end{equation}
having a spectral radius of $3.38$ and topological entropy $1.22$.  Advection
of material lines in this flow can be computed quickly using a map, rather
than solving the full flow numerically, therefore we can compute $L(t)$ for
larger $t$ and subsequently estimate the entropy more accurately at around
$1.48$. So the train-track bound is a reasonable $82\%$ sharp, to within a
percent.  This bound is slightly less sharp than for the examples shown in
Section~\ref{sec:numerics}. Perhaps this is to be expected since we did not
force the flow to execute a particular `good' braid.  Instead, we chose
periodic orbits that were easily identified analytically.  However, the bound
could easily be improved by including further periodic points in the
train-track calculation~\cite{Gouillart2006}. We leave this for a future
study.

\section{Discussion}
\label{sec:discussion}

We have shown how to compute lower bounds on the topological entropy of a
two-dimensional time- and spatially-periodic fluid flow using the `iterate and
guess' approach for constructing train-tracks.  For both one and two periodic
directions this approach may be applied directly by considering graphs
embedded on a cylinder or torus.  An alternative approach may be used for
singly-periodic flows, where a conformal transformation leads to a braid in an
annular domain with equal topological entropy.  In the latter approach,
performing the conformal mapping introduces an additional flow obstacle which
leads to a braiding motion with an additional, fictitious, stationary
stirrer. This confirms that in a spatially periodic flow it is possible to
create topological chaos with just two actual stirrers, rather than the three
required in a bounded or non-periodic infinite flow~\cite{Boyland2000}.  The
possibility of topological chaos also follows from the existence of
badly-ordered periodic orbits of the circle map associated with the stirrer
motion, which implies positive entropy~\cite{MacKay1984}.

Note that it does not follow that TC can be produced with a single stirrer on
a doubly periodic domain. The toroidal braid group with one stirrer is simply
$\{\tau_1,\rho_1\}$ (no $\sigma$'s). These two elements commute with each
other \cite{Birman1969}, so it is not possible to generate a non-trivial
braid. Intuitively, with only one stirrer there is nothing to wrap the stirrer
around and hence it is impossible to tangle fluid lines around anything. Flows
with one stirrer can of course still be chaotic, with a finite topological
entropy, but topological considerations with just one stirrer trajectory
reveal nothing.

The train-track approach actually provides more information than just the
growth rate of material lines. The entries in the transition matrix
eigenvector corresponding to the largest eigenvalue reveal where edges in the
train-track are accumulated at large times. In general the entries in this
vector are different from each other and account for the spatial inhomogeneity
in the relative number of striations between the stirrers, analogous to the
situation described by Alvarez {\it et al.}~\cite{Alvarez1998}.

In model Stokes flow simulations we have shown for the braids $\tau_1
\sigma_1$ and $\tau_1\sigma_1\rho_1^{-1}\sigma_1$ the computed entropy bounds
are around $90\%$ sharp. These bounds are surprisingly
good given that they are based only on the braid performed by the imposed
stirrer motions, however they could be made even sharper by including
information about other, secondary, periodic structures in the flow
\cite{Boyland1994,Gouillart2006}.

We found that the topological entropy of braids on the cylinder and torus was
quite high.  This is not surprising: the periodic boundary conditions offer
many more opportunities for complex braiding motions than the plane or a
bounded domain, thus causing material lines to grow rapidly.

Using the methods presented here, it was relatively simple to find braids in
the sine flow that have positive topological entropy.  The fact that this
topological entropy was~$82\%$ sharp when compared to the measured topological
entropy of the sine flow means that much of the observed chaos is embodied in
these orbits.  All these orbits are parabolic, which allows folding of
material lines around them---an important mechanism for chaos.

\section*{Acknowledgments}

Thanks to Toby Hall for help with train-tracks, to Igor Mezi\'c for helpful
discussions, and to an anonymous referee for clarifying several aspects of
Thurston--Nielsen theory.  This work was funded by the UK Engineering and
Physical Sciences Research Council grant GR/S72931/01.


\end{document}